\documentclass[aps,prx,twocolumn,superscriptaddress,showpacs,longbibliography]{revtex4-1}
\usepackage[colorlinks=true,citecolor=blue,linkcolor=blue,breaklinks=true]{hyperref}
\usepackage{amssymb}
\usepackage{graphicx}
\usepackage{amsmath}
\usepackage{mathtools}
\usepackage[export]{adjustbox}
\usepackage{epsfig}
\usepackage{times}
\usepackage{xcolor}
\usepackage{subfigure}
\usepackage{setspace}
\usepackage{bm}
\usepackage{calc}
\usepackage{natbib}
\usepackage[normalem]{ulem}
\usepackage{cancel}
\begin{document}

\definecolor{ao}{rgb}{0.0, 0.5, 0.0}
\newcommand{\acc}[1]{\textcolor{red}{\textit{[AC: #1]}}}
\newcommand{\ac}[1]{\textcolor{magenta}{\textit{ #1}}}
\newcommand{\acd}[1]{\textcolor{ao}{\textit{ #1}}}
\newcommand{\fm}[1]{\textcolor{black}{#1}}
\newcommand {\ba} {\ensuremath{b^\dagger}}
\newcommand {\Ma} {\ensuremath{M^\dagger}}
\newcommand {\psia} {\ensuremath{\psi^\dagger}}
\newcommand {\psita} {\ensuremath{\tilde{\psi}^\dagger}}
\newcommand{\lp} {\ensuremath{{\lambda '}}}
\newcommand{\A} {\ensuremath{{\bf A}}}
\newcommand{\Q} {\ensuremath{{\bf Q}}}
\newcommand{\kk} {\ensuremath{{\bf k}}}
\newcommand{\qq} {\ensuremath{{\bf q}}}
\newcommand{\kp} {\ensuremath{{\bf k'}}}
\newcommand{\rr} {\ensuremath{{\bf r}}}
\newcommand{\rp} {\ensuremath{{\bf r'}}}
\newcommand {\ep} {\ensuremath{\epsilon}}
\newcommand{\nbr} {\ensuremath{\langle ij \rangle}}
\newcommand {\no} {\nonumber}
\newcommand{\up} {\ensuremath{\uparrow}}
\newcommand{\dn} {\ensuremath{\downarrow}}
\newcommand{\rcol} {\textcolor{red}}

\newcommand{\fpc}[1]{\textcolor{blue}{\textit{[FP: #1]}}}
\newcommand{\fpm}[1]{\textcolor{black}{#1}}
\newcommand{\fpmm}[1]{\textcolor{black}{#1}}
\newcommand{\fps}[1]{\textcolor{black}{{#1}}}

\newcommand{\mpc}[1]{\textcolor{orange}{\textit{[MP: #1]}}}
\newcommand{\mpm}[1]{\textcolor{orange}{#1}}
\newcommand{\mps}[1]{\textcolor{orange}{\sout{#1}}}
\newcommand{\mpcancel}[1]{\textcolor{orange}{\cancel{#1}}}

\newcommand {\beq} {\begin{equation}}
\newcommand {\eeq} {\end{equation}}
\newcommand {\bqa} {\begin{eqnarray}}
\newcommand {\eqa} {\end{eqnarray}}


\begin{abstract}
How are \fpm{two-body} scattering and the resulting collective phenomena affected by preparing the \fpm{mediator of interactions} in different quantum states? 
This question has \fpm{recently} become experimentally relevant in a specific
non-relativistic version of QED implemented within materials, where standard techniques of quantum optics are available for the preparation of desired quantum states of the photon \fpm{mediating interactions between matter's constituents.}
We develop the necessary non-equilibrium approach for computing the
vertex function and find that, in addition to the energy and momentum structure of the scattering, a further structure emerges which reflects the Hilbert-space distribution of the \fpm{mediator's} quantum state. This emergent structure becomes non-trivial for non-Gaussian quantum states of the \fpm{mediator}, and can dramatically affect scattering and collective phenomena.
As a first application, we show that by preparing photons in pure Fock states one can enhance pair correlations, and even \fpmm{modify} the criticality of the superconducting phase transition. Our results also reveal that the thermal mixture of Fock states regularises the strong pair correlations present in each of its components, yielding the standard Bardeen-Cooper-Schrieffer criticality. \fpm{Besides the above QED platform, ultracold atomic mixtures are among the most promising candidates for the experimental implementation of these ideas.}

 \end{abstract}

  \title{Controlling Collective Phenomena Via the Quantum State of Interaction-Mediators:\\ Changing the \fpm{Criticality} of Photon-Mediated Superconductivity Via Fock States of Light}
 \author{Ahana Chakraborty}\email{ahana@physics.rutgers.edu}
 \affiliation{Department of Physics and Astronomy, Louisiana State University, Baton Rouge, Louisiana 70803, USA}
  \affiliation{Center for Materials Theory, Department of Physics and Astronomy, 
Rutgers University, Piscataway, NJ 08854, USA}
\author{Michele Pini}\email{michele.pini@uni-a.de}
\affiliation{Max Planck Institute for the Physics of Complex Systems, N\"othnitzer Str. 38, 01187, Dresden, Germany.}
\affiliation{Theoretical Physics III, Center for Electronic Correlations and Magnetism, Institute of Physics, University of Augsburg, 86135 Augsburg, Germany}
\author{Martina S. Z\"undel}
\affiliation{Max Planck Institute for the Physics of Complex Systems, N\"othnitzer Str. 38, 01187, Dresden, Germany.}
\affiliation{Univ. Grenoble Alpes, CNRS, LPMMC,
38000 Grenoble, France.}
\author{Francesco Piazza}\email{francesco.piazza@uni-a.de}
 \affiliation{Max Planck Institute for the Physics of Complex Systems, N\"othnitzer Str. 38, 01187, Dresden, Germany.}
 \affiliation{Theoretical Physics III, Center for Electronic Correlations and Magnetism, Institute of Physics, University of Augsburg, 86135 Augsburg, Germany}

\pacs{}
\date{\today}

\maketitle

\section{ Introduction }

Interactions between the constituents of matter are, at the
fundamental level, always described in terms of force carriers. In
the standard model of particle physics, these carriers (or mediators) are gauge
Bosons, the most commonly known being the photon, which mediates the
interaction between electric charges within quantum electrodynamics (QED).

The nature of the carrier controls the nature of the interactions between
the fermionic constituents of matter.
For instance, the mass, charge, or spin of the gauge Bosons
determine the properties of the force they mediate. Within the full
quantum mechanical description, the gauge Bosons are 
excitations of a field. Therefore, in addition to
the above single particle properties, a scattering event between two
fermionic particles will also in general depend on the quantum state
of the carrier gauge field. As an example, the electromagnetic field
can be found in a so-called coherent state, which follows the classical
Maxwell equations, or instead in a non-classical Fock state (also called
photon number state), which contains a non-fluctuating fixed number of
photons \cite{walls2007quantum}.

The following question can then be asked: How is the
scattering between the constituents of matter affected by preparing
the force carriers in different quantum states?
This work revolves around this issue, which is so far largely unexplored.
One reason for this lies in the fact that, in the context of high energy physics, a systematic experimental investigation is not feasible, as in particle colliders the gauge fields cannot be
prepared in a desired quantum state.
 \begin{figure}[]
   \label{fig:basic}
   \centering
   \includegraphics[width=\columnwidth]{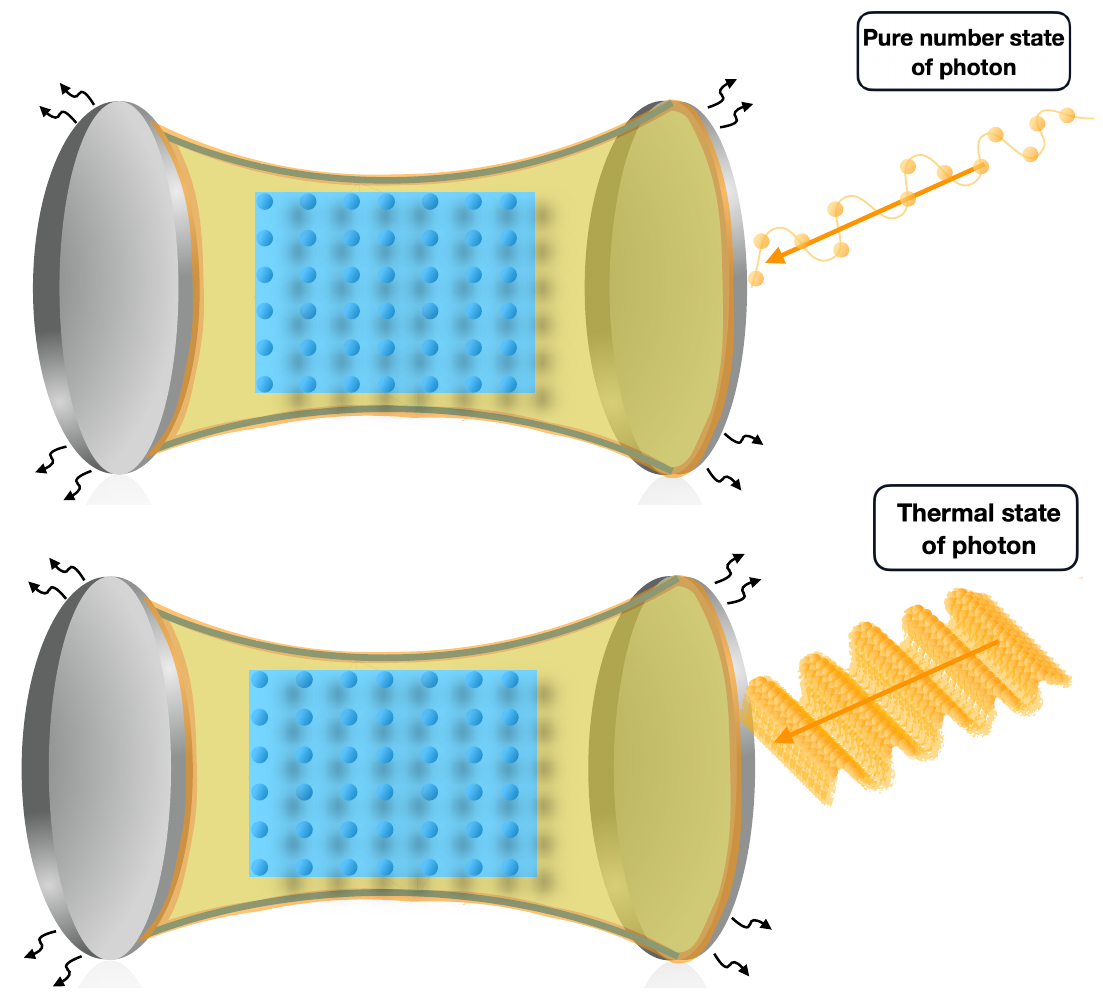}
   \caption{Sketch of a possible scenario: an ensemble of particles within a cavity interacts via photons which are prepared in a thermal equilibrium state or in a pure photon number (Fock) state.} 
 \end{figure}

On the other hand, the above question becomes 
relevant in a specific non-relativistic version of QED implemented
within materials \cite{mivehvar_2021,garcia2021manipulating,schlawin2021cavity}. Here, by confining the light around matter (for instance using mirrors forming a resonator), a selected number of
modes can be separated from the electromagnetic continuum, and standard
techniques of quantum optics become available for the preparation of
desired quantum states of photons. Importantly, mode confinement enhances
the coupling between photons and electrons, so that the material is
affected already by low-intensity electromagnetic fields, which are
more easily prepared in non-classical states. 
Recent experimental developments now allow to prepare non-Gaussian states of structured light \cite{rubinsztein-dunlop_roadmap_2016,forbes_quantum_2019,ra_non-gaussian_2020-1,erhard2020advances}, where the dimensionality (both in real and frequency space) of entanglement is enlarged by multiple modes beyond the binary polarization space.
Finally, the fact that
this regime of QED is realized within a material at finite electron densities, allows
to explore the effect of different quantum states of photons not only
on individual scattering events between electrons, but also on
emergent collective phenomena (see Figure \ref{fig:basic}).

  \begin{figure*}[t!]
   \centering
  \includegraphics[width=\textwidth]{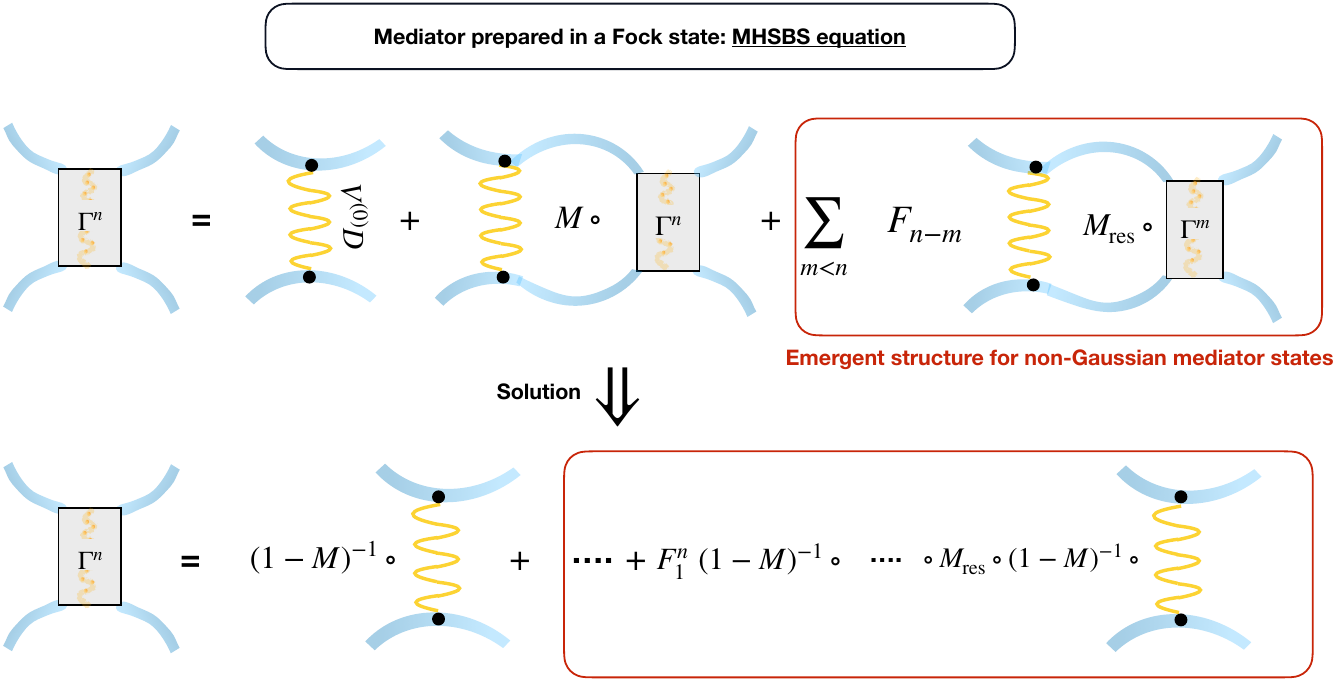}
  \caption{Emergent scattering structure in the \fpm{Mediator}-Hilbert-Space Bethe-Salpeter (MHSBS) equation (see Eq.~\eqref{eq:hierarchy_general}). Schematic representation of the equation for the scattering vertex function $\Gamma^{n}$ between two Fermions (blue lines) mediated by a single-mode \fpm{mediator} Boson (wiggled line) initially prepared in a pure Fock state  with occupation $n$. The properties of the Boson are encoded in the Green's function $D$, while the form factor of the Boson-Fermion coupling is $V^{(0)}$. Two Fermionic blue lines meet with the Bosonic wiggled line at the Yukawa vertex (given in Eq.~\eqref{Yukawa}) represented by black dots. The matrices $M$ and $M_{\rm res}$ encode the momentum, frequency, and causal structure of the scattering (with the symbol $\circ$ indicating the tensor product in all the corresponding spaces). In particular, $M_{\rm res}$ corresponds to scattering processes where a real mediator is fully absorbed/emitted by the Fermions. These processes are the ones depending on the quantum state of the mediator, and give rise to the emergent structure reflecting the Hilbert space of the latter. The vertex $\Gamma^{n}$ thus becomes coupled to vertices corresponding to smaller ($m<n$) Fock-space occupations $\Gamma^{m}$, with the coupling function $F_{n-m}$. The bottom row represents the form of the iterative solution of the MHSBS equation (see Eqs.~\eqref{eq:hier_iterative} and \eqref{gamma_n_itr}). 
 } 
   \label{fig:hierarchy}
   \end{figure*}
In order to answer our question of interest, an approach outside
the scope of equilibrium scattering and many-body theory is needed,
since the \fpm{mediator's} part of the system is externally prepared in an
arbitrary initial state. We achieve this by
developing a tailored real-time formulation of the Dyson equation for the two-particle vertex
function, known in thermal equilibrium as the Bethe-Salpeter equation \cite{abrikosov2012methods}. The advantage of our
approach is to reveal the difference between initial quantum states of
the \fpm{mediator} through an emergent structure of scattering
vertices. This structure, which is irrelevant when the \fpm{mediator} is prepared in a Gaussian state, adds to the usual energy-momentum and causal
structure of the scattering, and is determined by the wave function in the Hilbert-space of the \fpm{mediator}. Moreover, our method is amenable to
non-perturbative resummation techniques, allowing to describe collective phenomena, like instabilities towards macroscopically ordered phases.

 \begin{figure*}[t]
   \centering
   \includegraphics[width=0.70\textwidth]{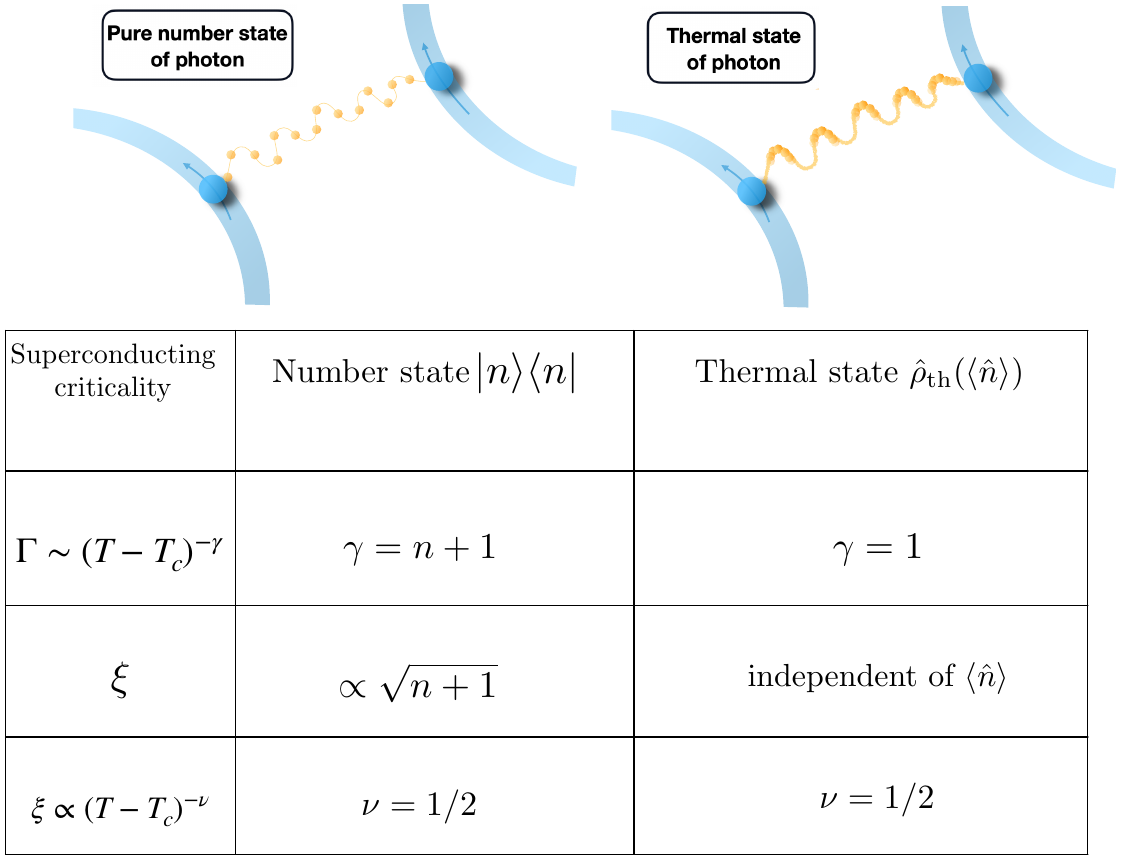}
   \caption{Superconducting critical behavior of electrons \fpm{for different quantum states} of the mediator photon in cavity QED. The \fpm{M}HSBS equation (see Fig.~\ref{fig:hierarchy}) is solved in the pairing channel at the superconducting critical point. Two quantum states of the \fpm{mediator} photons are compared: a pure Fock state with $n$ photons, and a thermal mixture (Eq.~\eqref{rho0_thermal}) with average number of photons $\langle\hat{n}\rangle=n_B(T)$.
   Different critical properties are considered: the susceptibility exponent $\gamma$ (see Section \ref{subsec:susceptibility}), the correlation length $\xi$ and its critical exponent $\nu$ (see Section \ref{subsec:correlation_length}). } 
     \label{table:result}
 \end{figure*}

After developing the general framework, we concentrate on a specific example in the
context of QED within materials: the case of
photon mediated electron pairing and superconductivity.
This recently proposed possibility provides a new scenario, with implementations
both in two-dimensional materials \cite{Schlawin2019,Gao2020} and ultracold atomic gases \cite{Colella_2019,Schlawin2019a,sheikan_2019}. 
In particular, by considering a specific tight-binding model of QED, recent studies have
found that, as an alternative to tuning the electromagnetic-mode energy, one
can project the photon onto number states to control the sign of the effective light-mediated, static interaction potential \cite{Sentef2020}, and thus
favour pairing and superconductivity over competing
instabilities \cite{li_manipulating_2020}. A modification of the effective light-matter coupling constant could
also be achieved by squeezing the photon field to enhance the proper
quadrature entering the electromagnetic vector field, as proposed in
the phonon case \cite{kennes_2017}.
On the other hand, in a previous work \cite{Chakraborty} we have shown that pairing can actually be dominated by
scattering processes which are not mediated by a static interaction
vertex as in the usual Bardeen-Cooper-Schrieffer (BCS) scenario, but
rather by resonant (non-adiabatic) photons.
Here we demonstrate that exactly those non-BCS processes directly depend
on the quantum state of the photon, and give thus rise to the emergent
structure of scattering vertices for non-Gaussian states. In
particular, by preparing the photons in pure Fock
states, one can enhance pair correlations in a controlled way, thereby
\fpmm{modifying} the criticality of the
superconducting phase transition. Our results also reveal
that the thermal mixture of Fock states regularizes the strong pair
fluctuations present in each of its components, yielding the usual BCS
criticality.

\fpmm{\section{Summary of the main results}}

\subsection{General result: qualitative modification of scattering through an emergent  structure reflecting the Hilbert space of a non-Gaussian mediator.}

\begin{enumerate}
    \item  For a generic form of the coupling between the fermionic
      matter and the bosonic mediator, \fpm{applicable to a broad class of systems including gauge theories, electron-phonon systems, and Bose-Fermi mixtures}, using a real-time formulation
      on the Keldysh contour we derive the equation for the two-particle scattering vertex between two Fermions with the
      bosonic mediator initially prepared in an arbitrary quantum
      state. This equation, which we name \fpm{Mediator}-Hilbert-Space
      Bethe-Salpeter (MHSBS) equation, is shown schematically in
      Fig.~\ref{fig:hierarchy} for the particular case of a single-mode mediator
      initially prepared in a pure Fock state. In the MHSBS equation,
      the scattering vertex acquires, in addition to the usual
      frequency, momentum and causal (Keldysh) components, further
      components labelled by Fock indices, i.e., representing the
      Hilbert space of the \fpm{mediator}. The couplings between the
      Hilbert-space components of the vertex are set by the initial
      quantum state of the \fpm{mediator}. 

      \fpmm{\item The additional Hilbert-space structure of the MHSBS equation is induced by resonant scattering processes involving the absorption/emission of a real (as opposed to virtual) mediator. Such processes are obviously dependent on the quantum state of the mediator.}
      
      \item \label{summary_hierarchy_gauss} When the state of the mediator is Gaussian,
      the couplings between the Hilbert-space components of the vertex are such that the emergent Hilbert-space structure in the MHSBS equation can be rearranged to become
      irrelevant, yielding the standard Bethe-Salpeter equation (see
      Fig.~\ref{fig:hierarchy_thermal}). The fact that the emergent
      structure becomes irrelevant does not mean that the Gaussian
      quantum state of the \fpm{interaction-mediator} plays no role, as it still affects the standard Bethe-Salpeter equation.
      
      \item \label{summary_hierarchy} In the particular case of a mediator prepared in a pure Fock state, the emergent scattering structure involves coupling between vertices corresponding to different Fock states with a number of \fpmm{mediator boson} smaller than the initial one. It can be solved in a hierarchical fashion starting from the vacuum state. Physically, this can be interpreted as the two Fermions scattering via the exchange of an ever decreasing number of photons from the initial one down to zero. \fpm{This very generally leads to a different functional dependence of the vertex function on external parameters, that is, to a qualitative different behavior of scattering and thus of collective phenomena.}
      
\end{enumerate}
    
    \subsection{Application to photon-mediated superconductivity: \fpmm{a different criticality for Fock states}.} 

  \begin{enumerate}
        
    \item We apply the \fpm{M}HSBS equation to the computation of the scattering vertex, for a specific QED setup belonging to the class described in the introduction, where cavity photons mediate interactions between electrons at finite density. 

    \item \fpmm{Photon-mediated interactions can induce the formation of electron pairs and ultimately a transition to a superconducting phase. We focus on the critical properties of this transition.} 

    \item We compare \fpm{the results for two different states of the photon (summarized in Fig.~\ref{table:result}):} i) a Gaussian thermal mixture (where the emergent scattering structure is irrelevant, see Fig.~\ref{fig:hierarchy_thermal}), and ii) a pure Fock state with $n$ photons (inducing a nontrivial emergent scattering structure of the type shown in Fig.~\ref{fig:hierarchy}).
    \fpmm{The preparation of Fock states of light within a cavity can be achieved for instance by injecting through the cavity mirrors a light pulse generated by one of the state-of-the-art sources of photon number-states. An alternative which avoids the in-coupling issues of the latter protocol is to prepare the Fock state of photons directly within the cavity, which can be done by using additional material's degrees of freedom to engineer the required optical nonlinearity.}

    \item The electron-electron vertex function in the pairing channel $\Gamma_{\rm pairing}$ diverges at the critical temperature \fpmm{for the superconducting transition. The corresponding critical exponent} $\gamma$ equals one in the standard BCS scenario where Bosons mediate simply a static attractive interaction. \fpmm{Instead, when the mediator boson (photon in our case) is prepared in a Fock state with $n$ particles, we find that $\gamma$ grows linearly with $n$. This is due to the emergent pair-scattering structure in the mediator Fock space, corresponding to a hierarchy of scattering processes where the electrons exchange an ever decreasing number of photons from the initial one down to zero (see point A.\ref{summary_hierarchy}). Indeed, this hierarchical structure leads to enhanced critical pair fluctuations, mathematically expressed by a photon-number-dependent power-law governing the divergence of the vertex function at the critical point (see Eq.\eqref{eq:n-dep_div_gamma}).}
   
    \fpmm{\item This photon-number-dependent power-law divergence of the vertex function also manifests in an enhanced electron pair correlation length $\xi$, which is increased by a factor $\sqrt{1+n}$.} 
    
    \item The enhanced critical pair fluctuations present when the photon is prepared in Fock states are regularized in the Gaussian thermal mixture, where the $\gamma$ and $\xi$ are independent of the photon number and agree with the BCS prediction. \fpmm{This regularization happens thanks to the peculiar Hilbert-space structure of the Gaussian thermal mixture, which combines the hierarchical structures of the scattering from each of its Fock-state component in such a way that they become irrelevant (see also point A.\ref{summary_hierarchy_gauss}).}
            
    \item \fpmm{The modification of the critical exponent appears in our calculations through} the modification of the effective dimensionality of the scattering space, which, as our MHSBS equation shows, is enlarged by the Hilbert space of the mediator.
    \fpmm{This very interesting finding shall stimulate future work where methods tailored to the computation of critical exponents (like the renormalization group analysis) will be extended to include the emergent Hilbert-space scattering-structure, in order to study possible changes in the universality class of the phase transition.}

\end{enumerate}

\section{Structure of the manuscript}

In Section \ref{sec:background} we provide a brief review of the background formalism. In Section \ref{BSIC} we derive our MHSBS equation. From Section \ref{LIS} on, we specify to the cavity QED model where photons mediate pairing between electrons, and compute the critical properties of the superconducting transition in Section \ref{critical}. In Section \ref{sec:polz}, we discussed justification of the time-independent approach we used to study the case of cavity-mediated superconductivity. Finally, in Section \ref{sec:conclusions} we provide concluding remarks and a brief outlook.

\section{Background Formalism}
\label{sec:background}

The approach employed in this work is based on
the Schwinger-Keldysh field theory (KFT), which is an action-based
formalism able to describe quantum many-body systems
out of equilibrium. The Keldysh partition function, $Z$ is defined
from the time-evolving density matrix $\hat{\rho}(t)$ as
$Z=Tr[U(\infty,0)\hat{\rho}(0)U(0,\infty)]$, where $U(t,0)$ and
$U(0,t)$ are the forward and backward time evolution operators,
respectively. These are represented by two path integrals, each
involving one of two
independent fields, $\chi_+(t)$ and $\chi_-(t)$
\cite{kamenevbook}. These two pieces of the path-integral contour are connected at $t=0$ by
the matrix element, $ \langle \chi_+(0) |\hat{\rho}(0) |  \chi_-(0)
\rangle$ of the density matrix of the system at the initial
time, as
\begin{equation}
Z=\int D[\chi_+,\chi_-] e^{\mathbf{i}\left[ S_{\mathrm{ev}} (\chi_+)- S_{\mathrm{ev}} (\chi_-)\right]} \langle \chi_+(0) |\hat{\rho}(0) |  \chi_-(0) \rangle.
\label{Z_generic}
\end{equation}
Here, $S_{\mathrm{ev}}$ is the action describing the time-evolution of the system over the forward and the backward contour. 

We start by considering a system of non-interacting bosonic/fermionic particles, described by the Hamiltonian
\begin{equation}
H_0= \sum \limits_{s} \omega_{s} ~b^{\dagger}_{s}~b_{s}.
\label{H_multi_diag}
\end{equation}
 Here $b^{\dagger}_{s}$ is the creation operator of the
 Bosons/Fermions in the single-particle state labeled by the index
 $s$, satisfying the usual commutation/anticommutation relations
 $(b_{s}b^{\dagger}_{s}\mp b^{\dagger}_{s}b_{s})=1$, with the
 convention where the upper/lower sign applies to the
 Bosons/Fermions. In this non-interacting case the action is simply
 \begin{eqnarray}
S_{\mathrm{ev}}(\chi_{\pm}) = \int_{-\infty}^\infty dt \sum \limits_s \chi^*_{\pm}(s,t) \left(  \mathbf{i} \partial_t -\omega_s  \right)  \chi_{\pm}(s,t).
\label{S_ev}
\end{eqnarray}

We will be interested in computing the Green's functions (GFs) of the
system. For this purpose we define the the generating functional,
which is obtained from $Z$ by adding sources $J_{\pm}(t)$ which couple
linearly to the fields $\chi_+$ and $\chi_-$ as \cite{kamenevbook},
\begin{eqnarray}
Z[J] &=& \int D[\chi_+, \chi_-] e^{\mathbf{i}\left[ S_{\mathrm{ev}} (\chi_+)- S_{\mathrm{ev}} (\chi_-)\right]} \langle \chi_+(0) |\hat{\rho}(0) |  \chi_-(0) \rangle \times  \nonumber \\
&&e^{\mathbf{i}  (\int \! dt \sum \limits_s   J_+^*(s,t) \chi_+(s,t)- \int \! dt  \sum \limits_s  J_-^*(s,t) \chi_-(s,t)+h.c.)}.~~~~~
\label{Z_J}
\end{eqnarray}

Due to the constraint imposed by the conservation of probability and
causality, not
all GFs are independent. It is thus useful to work in the rotated
basis of classical $\chi_{cl}=(\chi_++\chi_-)/\sqrt{2}$ and quantum
fields $\chi_{Q}=(\chi_+-\chi_-)/\sqrt{2}$. For instance, among the
one-particle GFs, only two out of four are independent, namely the 
retarded and the Keldysh (also called statistical) GF. Those are defined as,
\begin{eqnarray}
G_R(s,t,t')&=&-\mathbf{i}\langle \chi_{cl}(s,t) \chi_Q^*(s,t') \rangle= \frac{\mathbf{i}\partial^2 Z[J]}{\partial J^*_{Q}(s,t) \partial J_{cl}(s,t')} \nonumber \\
G_K(s,t,t')&=&-\mathbf{i}\langle \chi_{cl}(s,t) \chi_{cl}^*(s,t') \rangle=\frac{\mathbf{i} \partial^2 Z[J]}{\partial J^*_{Q}(s,t) \partial J_{Q}(s,t')}, \nonumber \\
\end{eqnarray}
where $J_{cl,Q}=(J_+ \pm J_-)/\sqrt{2}$ are the classical and quantum component of the linear sources. As the nomenclature suggests, $G_R$ represents the retarded evolution of the system governed by the dispersion (and the lifetime) of the modes, and hence does not depend on the initial conditions. On the other hand, $G_K$ explicitly depends on the initial conditions carrying their memory during the time-evolution.

In order to proceed and derive equations of motions for the
GFs, we have to deal with the matrix element of
the initial state. As illustrated in the next two Sections, we will do
this by exponentiating the matrix element of the initial state, so that $Z$ can be expressed as a path-integral with an action which incorporates both the information about the initial conditions as well as the time-evolution under the Hamiltonian

\subsection{Single-particle Green's functions for thermal initial states}
  In this sub-Section, we will illustrate how to incorporate the
  matrix element of the initial state in the action for a thermal initial condition and obtain the GFs.
 We consider a Gaussian thermal initial density matrix at temperature $T=1/\beta$ given by, 
 \begin{equation}
  \hat{\rho}(0)=\hat{\rho}_{\rm th}=\frac{\sum \limits_{\{n_s\}} e^{-\beta \sum \limits_{s} \omega_s n_s} |\{n_s\} \rangle \langle \{ n_s \}|}{\sum \limits_{\{n_s\}} e^{-\beta \sum \limits_{s} \omega_s n_s}} ,
  \label{rho0_thermal}
\end{equation}
with $| \{ n_{s} \} \rangle = | n_1,n_2,...\rangle$ being a Fock state
with $n_1$ particles in the mode $1$, $n_2$ in the mode $2$, etc..
In this case, the matrix element of the initial state can be readily exponentiated as a bilinear term of the initial fields as \cite{kamenevbook},
\begin{equation}
  \langle \chi_+(0) |\hat{\rho}(0) |  \chi_-(0) \rangle =  \prod_s(1 \mp \rho_s)^{\mp 1}~e^{~ \pm\sum \limits_s \rho_{s} \chi^*_+(s,0) \chi_-(s,0)},
 \end{equation}
  where $\rho_s=\exp(-\beta \omega_s )$. Hence, the information about
  the thermal initial conditions can be easily added as a quadratic
  (bi-linear) term in the action as
  \begin{eqnarray}  
Z &=&\prod_s(1 \mp \rho_s)^{\mp 1} \, \times \nonumber \\
&&\int \! D[\chi_+,\chi_-] e^{\mathbf{i}\left[ S_{\mathrm{ev}}
    (\chi_+)- S_{\mathrm{ev}} (\chi_-)+  \delta S(\chi^*_+,\chi_-;
    \rho_s) \right]}, \nonumber 
        \label{eq:partition_thermal}
\end{eqnarray}
with
\begin{equation}
  \delta S(\chi^*_+,\chi_-;\rho_s)= \mp\mathbf{i} \sum \limits_s \rho_s \chi^*_+(s,0) \chi_-(s,0).
  \label{DeltaS_thermal}
\end{equation}  
This makes systems starting from Gaussian thermal initial density matrices amenable to the standard KFT techniques  \cite{kamenevbook}. 
For non-interacting systems (evolving with the Hamiltonian given in
Eq. \eqref{H_multi_diag}), the whole action is
quadratic and the functional integrals over the fields can be
performed exactly to yield the one-particle GFs,
\begin{eqnarray}
G_R(s,t,t')&=&-\mathbf{i} \Theta(t-t') e^{-\mathbf{i}\omega_s (t-t')}, \nonumber \\
G_K(s,t,t')&=&-\mathbf{i} \frac{1\pm\rho_s}{1\mp\rho_s} e^{-\mathbf{i}\omega_s (t-t')},
\label{G_thermal}
\end{eqnarray}
where $(1\pm\rho_s)/(1\mp\rho_s)=1\pm2 n_{B(F)}$ is related to the
Bose/Fermi distribution $n_{B(F)}$ in thermal equilibrium.

\subsection{Single-particle Green's functions with arbitrary
  initial states}
\label{sec:spGF_nonthermal}

  For an
  arbitrary $\hat{\rho}(0) $, exponentiation of its matrix element in Eq.~\eqref{Z_generic} can not be performed
  to obtain a $\delta S$ containing only bi-linear terms involving
  the initial fields $\chi_+(0)$ and $ \chi_-(0)$. Hence, even
  non-interacting systems starting from non-thermal states $
  \hat{\rho}(0)  $ yield a non-Gaussian field theory which cannot be
  treated exactly within the standard KFT. Recently, in
  Ref.~\cite{Chakraborty2},  an efficient prescription has been proposed to exponentiate $  \hat{\rho}(0)  $ within the action-based KFT formalism, which we will briefly review here.

To illustrate this, we first assume that the system is initialized in
a pure Fock state $| \{ n_{s} \} \rangle = | n_1,n_2,...\rangle$,
 \begin{equation}
 \hat{\rho}(0) = | \{ n_{s} \} \rangle \langle \{ n_{s} \} |.
 \label{rho0_Fock}
 \end{equation}
In this case, $\langle \chi_+(0) |\hat{\rho}(0) |  \chi_-(0) \rangle$ can be exponentiated by introducing a quadratic source $u_s$ which couples to the bi-linears of the initial fields $\chi^*_+(s,0) \chi_-(s,0)$ for all the modes $s$, and finally taking appropriate derivatives w.r.t.~$u_s$. For the state given in Eq.~\eqref{rho0_Fock}, the matrix
element can be written as
 \begin{equation}
 \langle \chi_+(0) |\hat{\rho}(0) |  \chi_-(0) \rangle \! =\mathcal{L} ~e^{\pm \sum \limits_{s} u_s \chi^*_+(s,0) \chi_-(s,0)} \Bigg|_{u_s=0},
 \label{exponentiation}
 \end{equation}
 where the derivative operator,
 \begin{equation}
 \mathcal{L}=\prod \limits_{s} \! \frac{1}{n_s!} \left(\!  \frac{\partial}{\partial u_s} \!\right)^{\!n_s}
 \end{equation}
 encodes the information about the initial Fock state.
 
 The corresponding partition function of the system 
can be thus written as 
\begin{eqnarray}
Z\!=\mathcal{L} ~\left[ \int D[\chi_+,\chi_-] e^{\mathbf{i}\left[ S_{\mathrm{ev}} (\chi_+)- S_{\mathrm{ev}} (\chi_-)+  \delta S(\chi^*_+,\chi_-;u) \right]}\right] \bigg|_{u_s=0}, \nonumber 
\end{eqnarray}
with
\begin{equation}
\delta S(\chi^*_+,\chi_-;u) =\mp\mathbf{i} \sum_s u_s \chi^*_+(s,0) \chi_-(s,0).
\label{DeltaS_nonthermal}
\end{equation}

The prescription for computing the GFs is then organized in two
steps.

(i) Define an intermediate $u_s$-dependent partition function in presence of the linear, $J_{\pm}$, and quadratic, $u_s$ sources,
\begin{eqnarray}
Z[J;u]&=&\! \! \int \! \!   D[\chi_+,\chi_-] e^{\mathbf{i}\left[ S_{\mathrm{ev}} (\chi_+)- S_{\mathrm{ev}} (\chi_-)+  \delta S(\chi^*_+,\chi_-;u) \right]} \times \nonumber \\
&&e^{\mathbf{i}  (\int \! dt \sum \limits_s J_+^*(s,t) \chi_+(s,t)- \int \! dt \sum \limits_q  J_-^*(s,t) \chi_-(s,t)+h.c.)}.
\end{eqnarray}
By taking derivatives of $Z[J;u]$ w.r.t. the linear sources we obtain an intermediate $\{u_s\}$-dependent Green's functions $G(s,t,t';u)$ as,
\begin{equation}
\frac{\mathbf{i}\partial^2 Z[J;u]}{\partial J^*(s,t) \partial J(s.t')}=\left[ \prod \limits_s (1\mp u_s)^{\mp 1} \right] G(s,t,t';u),
\end{equation}
where we omitted the contour indices for simplicity.
 At this point it is important to note that the normalization of the intermediate partition function $Z[J=0;u]=\prod_s (1\mp u_s)^{\mp 1}$ crucially depends on the initial sources $\{u_s\}$ and hence we need to keep it explicitly in the path integral (unlike the Gaussian thermal case where $Z[J=0]=1$). The intermediate retarded and Keldysh Green's functions take the form,
 \begin{eqnarray}
 G_R(s,t,t';u)&=& -\mathbf{i} \Theta(t-t') e^{-\mathbf{i}\omega_s (t-t')} \nonumber \\
  G_K(s,t,t';u)&=& -\mathbf{i} \frac{1\pm u_s}{1\mp u_s} e^{-\mathbf{i}\omega_s (t-t')},
 \end{eqnarray}
 where the retarded component is (as expected) independent of the
 initial sources and hence the information about the initial
 conditions of the system is solely carried by the Keldysh component.
 
(ii) Take the derivatives of $G(s,t,t';u)$ w.r.t. $\{u_s\}$ to calculate the physical GFs ${G}(s,t,t')$ of the system
. This yields
 \begin{eqnarray}
 {G}_R(s,t,t')&=&  \mathcal{L} \left[\prod \limits_s(1\mp u_s)^{\mp 1}    G_R(s,t,t';u) \right] \Bigg|_{ u_s =0}\nonumber \\
 &=& -\mathbf{i} \Theta(t-t') e^{-\mathbf{i}\omega_s (t-t')} \nonumber \\
 {G}_K(s,t,t')&=& \mathcal{L} \left[  \prod \limits_s(1\mp u_s)^{\mp 1}  G_K(s,t,t';u) \right] \Bigg|_{ u_s =0}  \nonumber \\
 &=& -\mathbf{i} \left(1\pm 2 n_s \right) e^{-\mathbf{i}\omega_s (t-t')}.
 \label{Green_physical_Fock}
 \end{eqnarray}
 Here, for the sake of simplicity of the notation, we use $G$ with an explicit $u$ dependence in the argument as an intermediate GF, while the one without such dependence is the physical GF.

 We note that interchanging the order of taking derivatives w.r.t. $J$
and $u$ is permissible, as $J$ are linear sources and $u$
quadratic sources. This is crucial for the major simplifications
allowed by this method \cite{Chakraborty2} (see also next Section).

An immediate extension of the above formalism for initial pure Fock
states is to include a density matrix which is diagonal in the Fock basis,
\begin{equation}
 \hat{\rho}(0) =\sum \limits_{\{n_s\}} c_{\{  n_{s}\}} | \{ n_{s} \} \rangle \langle \{ n_{s} \} |,
 \label{rho0_diagonal}
\end{equation}
where $\sum_{\{n_s\}} c_{\{  n_{s}\}}=1$. In this case the derivative operator which encodes the information about $ \hat{\rho}(0)$ takes the form 
\begin{equation}
\mathcal{L}=\sum_{\{n_s\}} c_{\{  n_{s}\}}\prod_{s} \frac{1}{n_s!} \left( \frac{ \partial}{\partial u_s }\right)^{n_s}.
\label{L_diagonal}
\end{equation}
As before, applying $\mathcal{L}$ on the intermediate $\{u_s\}$-dependent Green's functions $G(s,t,t';u)$ (together with the $\{u_s\}$-dependent normalization) we obtain the final physical Keldysh GF,
\begin{eqnarray}
G_K(s,t,t')&=&-\mathbf{i} \sum \limits_{\{ n_s \}} c_{\{ n_s \}}\left(1+2 n_s \right) e^{-\mathbf{i}\omega_s (t-t')} ,\nonumber \\
\end{eqnarray}
while the retarded component of the correlation function remains unaffected by the initial condition and is given by the first two lines of Eq.~\eqref{Green_physical_Fock}.

We note that in both the examples of initial density matrices given above in Eq.~\eqref{rho0_Fock} and \eqref{rho0_diagonal}, the dynamics of the system is invariant under time-translation as the Hamiltonian is diagonal in the same Fock basis. 
The time-translation invariance is broken when we consider a generic initial density matrix of the multi-mode system which also has non-diagonal elements expanded in the Fock basis. Within this extended KFT formalism, we can deal with a generic number-conserving $\hat{\rho}(0)$ of the form,
\begin{equation}
 \hat{\rho}(0) =\sum \limits_{\{n_s\},\{ m_s\}} c_{\{  n_{s}\}\{m_s\}} | \{ m_{s} \} \rangle \langle \{ n_{s} \} |,
 \label{rho0_generic}
\end{equation}
with the constraint $\sum_s n_s=\sum_s m_s$. In this case the physical
Keldysh GF becomes a function of two time variables:
\begin{equation}
G_K(s,s',t,t')= -\mathbf{i} \left(1+2\langle b^{\dagger}_s b_{s'}   \rangle_0 \right) e^{-\mathbf{i}(\omega_s t- \omega_{s'}t')},
\end{equation}
while the retarded GF is unaffected by the initial condition and
thus remains time-translation invariant. 
In this case, the system exhibits a non-trivial time-evolution induced by the memory of the initial state.

 \section{Vertex function for interaction mediator initialized in arbitrary quantum states} \label{BSIC}
 
 After having reviewed how to compute single-particle GF in a
 system of Bosons or Fermions initialized in a non-Gaussian, non-thermal state, we now turn to the case
 of interest, which is the one where Fermions (the constituents of
 matter) interact with Bosons (the \fpm{mediators}) via a Yukawa type
 coupling, and the Bosons are initialized in a
non-thermal state.
We will be interested in studying how Fermion-Fermion scattering mediated by the Boson, and
ultimately collective phenomena,
are affected by the state in which the Boson is initialized.
This requires us to develop an extension of the standard Dyson
equation for the two-particle vertex function (also called Bethe Salpeter
equation), allowing to deal with the memory of the non-thermal, non-Gaussian initial
conditions of the Boson.

The fermionic matter is initialized in a thermal density matrix given by Eq.~\eqref{rho0_thermal}. The contribution to the non-interacting part
of the action coming from the Fermions is $S_F(\psi_{cl},\psi_{Q})$ defined in terms of the Grassmann fields $\psi_{cl},\psi_{Q}$. This takes the thermal form composed of Eqs.~\eqref{S_ev} and
\eqref{DeltaS_thermal}. By inverting $S_F$ we can obtain the non-interacting
fermionic GFs given in Eqs.~\eqref{G_thermal}. On the other hand, the
Bosons are initialized to the generic diagonal density matrix given in
Eq.~\eqref{rho0_diagonal}, which covers a broad class of initial conditions. Differently from the fermionic counterpart, the non-interacting
bosonic action has the contribution given by
Eq.~\eqref{DeltaS_nonthermal}, which depends on the initial quadratic
source $u$. It is useful to express the bosonic action
$S_{B}(X_{cl},X_Q;u)$ in terms of real position (also called
quadrature) fields $X_{cl,Q}(s,t)=\big[\chi_{cl,Q}(s,t)+\chi^{*}_{cl,Q}(s,t)\big]/\sqrt{2\omega_s}$, yielding the non-interacting
$u-$dependent (intermediate) bosonic GFs of the form,
\begin{eqnarray}
D_K(s,t,t';u)
=-\mathbf{i} \frac{1+u_s}{1-u_s} \frac{\cos(\omega_s (t-t'))}{2\omega_s}, \nonumber \\
D_R(s,t,t')
=-\Theta(t-t')  \frac{\sin(\omega_s (t-t'))}{2\omega_s}.
 \label{DK_u}
\end{eqnarray}

We consider the following form of the Yukawa-type coupling
 \begin{equation}
 H_{\rm int} = \sum \limits_{k,k',s} \sqrt{2\omega_s}  g_{k,k',s} c^{\dagger}_{k'} c_{k}X(s,t),
 \label{Yukawa}
 \end{equation}
 where $c^{\dagger}_k$ ($c_k$) is the creation (annihilation)  operator of the Fermions with momentum $k$ and $X(s,t)$ is the position operator of the Bosons in the $s$ mode at time $t$. \fpm{We stress that the form of coupling \eqref{Yukawa} can describe a broad variety of cases, including the coupling of matter to vector potentials like in gauge theories, the coupling between electrons and phonons, or the coupling between the Bose-condensed component and the fermionic component of a binary mixture, as realized both in ultracold-atomic \cite{schreck_quasipure_2001,hadzibabic_two-species_2002} and in exciton-electron \cite{tan_interacting_2020,cotlet_2016} systems}.
 The corresponding part of the
 action reads
 \begin{eqnarray}
 &&S_{\rm int}=-\int \limits_{-\infty}^{\infty} dt \sum \limits_{ k,k',s}g_{k,k',s}\sqrt{2\omega_s} \times \nonumber \\
 &&  \Big[
\{ \psi^*_{cl}(k',t) \psi_{Q}(k,t)+ cl \leftrightarrow Q  \}X_{cl}(s,t) \nonumber \\
&&+
 \{ \psi^*_{cl}(k',t) \psi_{cl}(k,t)+ cl \leftrightarrow Q \}X_{Q}(s,t)
\Big]. \nonumber \\
\label{S_lightmatter}
 \end{eqnarray}
After integrating out the bosonic degrees of freedom (which can be
done exactly), we obtain the intermediate partition function $Z[u]$ involving only fermionic fields of the form,
\begin{equation}
 Z[u]=\frac{1}{\prod \limits_{s}(1-u_{s})}\int D[\psi_{cl,Q}]
 e^{\mathbf{i}S_F(\psi_{cl},\psi_Q)+\mathbf{i}S_{\rm eff}(\psi_{cl},\psi_Q;u)},
\end{equation} 
 where the four-Fermion effective interaction term mediated by the Bosons is given by a sum of terms in Keldysh space with the following structure,
 \begin{eqnarray}
 S_{\rm eff}(\psi_{cl},\psi_Q;u) &=&- \int dt dt'  \! \!\! \sum \limits_{ \{k_i\},s} V^{(0)}(\{k_i\},s) D(s,t,t';u) \times \nonumber \\ && \psi^*(k_1,t) \psi^*(k_2,t') \psi(k_3,t') \psi(k_4,t).~~~~~~~
 \label{S_eff}
 \end{eqnarray}

 The strength of the quartic interaction is $ V^{(0)}(\{k_i\},s)
=\omega_s~ g_{k_3,k_4,s}~g^*_{k_1,k_2,s}$.
Here we have omitted all Keldysh indices for the sake of simplicity,
since the results of the present section can be illustrated without explicit reference to the
Keldysh structure. This does not mean the latter is not important, as
the memory of the initial quantum state in which the \fpm{mediator} Boson is
prepared is encoded only in the Keldysh component of $D(s,t,t';u)$. 
Due to the causality structure of 
the theory, each component of $D(s,t,t';u)$ can appear only together
with certain combinations of the 
$(cl,Q)$ indices of the Fermion fields. We refer to Appendix \ref{app:action} for a detailed
discussion.

\begin{figure*}[t!]
   \centering
  \includegraphics[width=\textwidth]{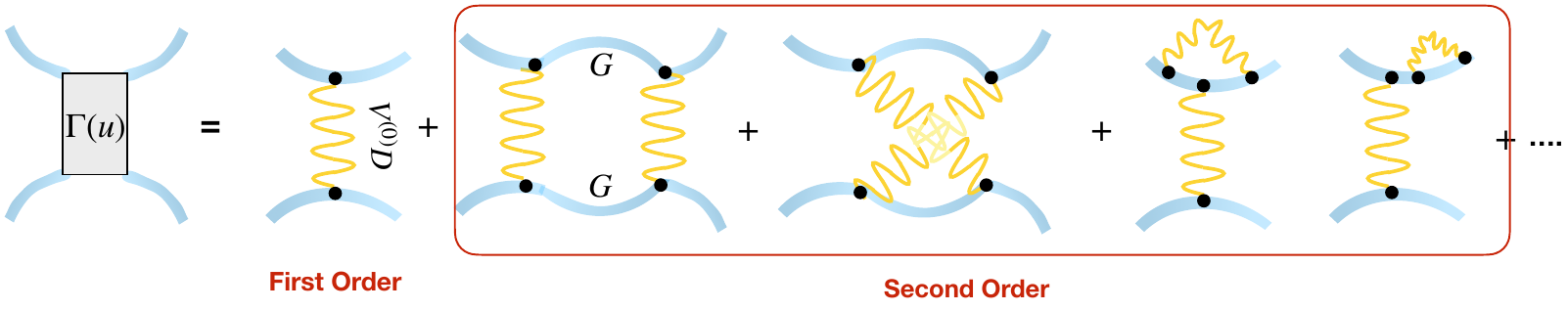}
  \caption{The perturbative expansion of the intermediate vertex function $\Gamma(u)$, which takes the same form as in the thermal- equilibrium case \cite{abrikosov2012methods}. The diagrams are arranged in powers of the coupling factor
$V^{(0)}(\{k_i\},s)$, and are shown up to second order. Fermions are represented by solid blue lines ($G$ is the non-interacting Fermionic Green's function) while the \fpm{mediator} Bosons are represented by yellow wiggled lines. At first order, $\Gamma(u)$ is simply the bare interaction vertex $\Gamma(u)=V^{(0)}(\{k_i\},s) D(s,t,t';u)$. Similarly, in all higher order diagrams, the $u_s$ dependence is solely coming through the \fpm{mediator} GF $D(s,t,t';u)$, and the contribution from the \fpm{mediator's} initial state in each diagram can be computed analytically.}
   \label{fig:vertex_pert}
   \end{figure*}

The physical partition function of the system is obtained by
taking the appropriate set of derivatives of $Z[u]$ w.r.t.~$\{u_s\}$ through
the operator $\mathcal{L}$. The detailed form of the operator
$\mathcal{L}$ depends the specific form of $\hat{\rho}(0)$ of the Bosons. For the particular case
considered in this Section in Eq.~\eqref{rho0_diagonal},
$\mathcal{L}$ is given by Eq.~\eqref{L_diagonal}. As illustrated in the previous Section, to calculate the
physical GFs of the interacting Fermions, we add
linear sources $J(k,t)$ coupled to the fields
$\psi(k,t)$ at all times $t$,
and then take derivatives of the physical generating functional $Z[J]$
w.r.t.~$J$ to obtain the $m-$particle physical GF
${G}^{F}_m(k_1,...,k_m,t_1,,...t_m)$.
The non-linearity of the derivative operator $\mathcal{L}$, together with the presence
of the crucial $u_s$-dependent normalization $1/\prod_{s}(1-u_s)$ of
the partition function, makes several useful properties of the theory, which could be exploited for Gaussian thermal initial conditions, inapplicable here. A major complication comes from the fact that Wick's theorem cannot be used to factorize the physical higher-order GFs in terms of two-point GFs when the action is Gaussian \cite{Chakraborty2}.
This leads to the breakdown of standard
perturbative and non-perturbative approximations, like the Dyson or
Bethe-Salpeter equations normally used for systems
initialized in a Gaussian thermal state \cite{abrikosov2012methods}.
In order to overcome these problems in our specific computation of the
Fermion-Fermion vertex function with a force \fpm{mediator} prepared in a
non Gaussian initial state, we adopt a two-step procedure like the one used for
the case of purely bosonic/fermionic model of Section
\ref{sec:spGF_nonthermal}. After having integrated out the \fpm{mediator}, we are dealing with a purely fermionic action. 
We first define an intermediate $\{u_s\}$-dependent GF
   ${G}^F_m(k_1...,k_{2m},t_1,...t_{2m};u)$ obtained from
   $Z[J,u]$ by taking derivatives with respect to the $J$'s, and
   subsequently take the final derivatives w.r.t.~$\{u_s\}$ via the operator
   $\mathcal{L}$ to obtain the
   physical GF ${G}^F_m(k_1...,k_{2m},t_1,...t_{2m})$.
   Even though we employ the same procedure, the structure of the
   resulting equations in our case is very different from the one
   obtained for a purely fermionic theory where the interactions are not mediated by
   a Bosons and the Fermions themselves are initialized in a
   non-Gaussian state. The reason is that in our case the operator
   $\mathcal{L}$ acts on the interaction part of the action $S_{\rm
     eff}$ as it contains the \fpm{mediator} GF, while in the purely
   fermionic case $\mathcal{L}$ acts only on the quadratic part
   $\delta S$.
   Although taking derivatives w.r.t.~$\{u_s\}$ and obtaining a closed expression
 for ${G}^{F}_m$ can be
 non-trivial, the above two-step construction makes the problem
 manageable in the best possible way: at the intermediate stage we can
 exploit all the useful properties also available in the Gaussian thermal case
 for approximating the equations for the GFs, and in particular diagrammatic
 perturbative approaches and resummation techniques to obtain the intermediate ${G}^{F}_m(u)$.

The next step is now to derive the equation for the vertex
function. We will focus on two-particle vertices, and start from
 the two-particle physical GF, $G^{F}_2(k_1,..k_4,t_1,..t_4)=
 \mathcal{L}[G^{F}_2(k_1,..k_4,t_1,..t_4;u)/\prod_{s}(1-u_s)]|_{u_s=0}$. The
 intermediate $G^{F}_2(k_1,..k_4,t_1,..t_4;u)$ can be formally constructed as a perturbation expansion in terms of the intermediate two particle vertex function $\Gamma(k_1,..k_4,t_1,..t_4;u)$ defined as,
 \begin{widetext}
 \begin{equation}
 G^{F}_2(k_1,..k_4,t_1,..t_4;u)=\int \prod \limits_{i} dt_i G(k_1,t_1,t'_1)G(k_2,t_2,t'_2)G(k_3,t_3,t'_3)G(k_4,t_4,t'_4) \Gamma(k_1,..k_4,t'_1,..t'_4;u).
 \label{Gamma_int}
 \end{equation}
 \end{widetext}
 Here the external points at times $t_i$ are connected to the internal
 points at time $t'_i$ by non-interacting Fermion GFs $G$
 (independent of $\{u_s\}$), and the effects of the Fermion-Fermion
 interactions are included in
 the definition of $\Gamma$, which thus solely carries the
 $\{u_s\}$ dependence \footnote{Let us point out that the
   definition of $\Gamma$ in Eq.\eqref{Gamma_int} is different from that
   in the standard thermal field theory where the external Fermion
   lines are full interacting Fermion Green's functions.}.
 The physical two-particle vertex function is then given by,
 \begin{equation}
 \Gamma(k_1,..k_4,t_1,..t_4)= \mathcal{L} \left[ \frac{ \Gamma(k_1,..k_4,t_1,..t_4;u)}{\prod \limits_s (1-u_s)} \right]\Bigg |_{u_s=0}.
  \label{Gamma_ph}
 \end{equation}
This function contains all the information about the 
initial state of the \fpm{mediator}, and thus gives a
self-contained quantity to investigate how the latter affects the Fermion-Fermion scattering. Moreover, a
divergence in the physical vertex function signals an instability
towards a macroscopically ordered phase where the
corresponding fermionic bilinear acquires a finite expectation value. We will consider the example of pairing and superconductivity
in Section \ref{LIS}.

Now we turn our attention to the construction of a diagrammatic expansion of
the intermediate vertex function $\Gamma(u)$. The simplest one is a
perturbative expansion in powers of the quartic interaction potential
$V^{(0)}(\{k_i\},s)$ as shown in
Fig.~\ref{fig:vertex_pert}. Each Boson
line corresponds to $D(s,t,t';u)$ and carries the $\{u_s\}$
dependence coming from the initial state. Hence, if
we truncate the perturbation series at certain order (say $n$, see
Fig.~\ref{fig:vertex_pert}), the $u_s$ dependence of the a given diagram has an
analytic form $(1+u_s)^p/(1- u_s)^{p+1}$ (including the
normalization) with $p \leq n$, where the equality holds for the
  diagram where each \fpm{mediator's} GF is a Keldsyh GF. This means
that the action of the derivative operator $\mathcal{L}$ can be computed exactly to obtain a closed-form analytical answer for the physical vertex $\Gamma$.
\fpm{Here, we are assuming the perturbation series of $\Gamma(u)$ to be convergent for $u_s \to 0$, i.e.~the convergence radius $R_s$ of the series with respect to the variable $z_s=(1+u_s)/(1-u_s)$ is larger than 1. The requirement $R_s>1$ comes from the fact that the physical vertex (\ref{Gamma_ph}) is evaluated at $|z_s|=1$ ($u_s=0$). The possible scenario where this assumption is broken (see e.g. \cite{brando_review}), which would lead to
an infrared catastrophe with a non-analytic dependence of $\Gamma(u)$
on $u_s$ at $u_s \to 0$, is not considered here.}

A perturbative computation of $\Gamma$ is however not sufficient
to capture collective phenomena in many cases, the most prominent one
being the description of instabilities towards macroscopically ordered
phases. The required non-perturbative closed-form
equation
can be derived for the intermediate vertex $\Gamma(u)$.
We demonstrate this using a widely used resummation technique for a
selected class of diagrams, called ladder resummation {(corresponding to the leftmost of the second-order diagrams in Fig. \ref{fig:vertex_pert})}, applied to $\Gamma(u)$: 
\begin{eqnarray}
\Gamma(u)&=& V^{(0)} D(u) + V^{(0)} D(u) \circ G  \circ G \circ \Gamma(u) \nonumber \\
\label{Gamma_ladder}
\end{eqnarray}
 where $\circ$ stands for the appropriate integration or summation of
 the internal indices in position, time, and Keldysh space, and
 $V^{(0)} D(u)$ is the bare vertex function in the infinite series.
 The above equation can be rewritten as,
\begin{eqnarray}
\Gamma(u)
&=& V^{(0)} D(u) +  M_{\rm nonres} \circ \Gamma(u) + f(u) M_{\mathrm{res}} \circ \Gamma(u). \nonumber \\
\label{Gamma_ladder_separated}
\end{eqnarray} 
 Here, we have separated the $u$-independent contribution (2nd term in R.H.S.), containing $D_{R,A}$, from the $u$-dependent contribution,  containing $D_K(u)$. The full $u$-dependence of the second term Eq.~\eqref{Gamma_ladder} is collected in the function $f(u)$. The rest (including $V^{(0)},G$ and $D(u=0)$) has been absorbed into the tensors $M_{\rm nonres}$ and 
 $M_{\mathrm{res}}$. The subscript of $M_{\mathrm{res}}$ here refers to resonant processes involving the absorption/emission of a real (as opposed to virtual) \fpm{mediator}. It is physically clear that it is those processes that depend on the quantum state of the \fpm{mediator}. Here the ladder is constructed out of the non-interacting fermionic GFs, $G$, which will simplify the computation of the physical vertex as the $\{u_s\}$-derivatives will act only on the \fpm{mediator's} GFs. This type of ladder resummation scheme is the standard approximation applied in thermal equilibrium to study instabilities associated with a divergence of $\Gamma$ \cite{abrikosov2012methods}.

\subsection{\fpm{Mediator} initialized in a pure Fock state}

\begin{figure*}[t!]
   \centering
  \includegraphics[width=\textwidth]{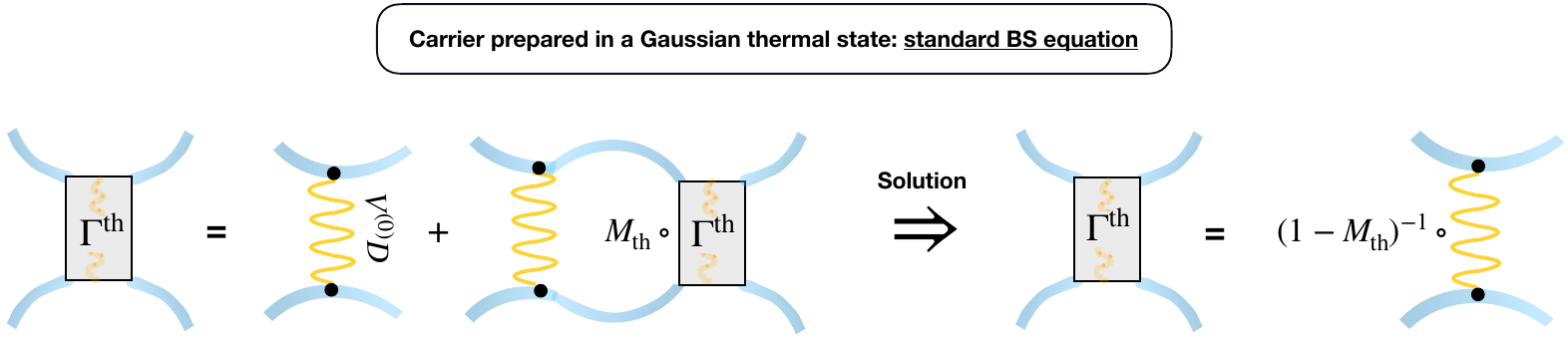}
  \caption{Scattering structure in the standard Bethe-Salpeter (BS) equation (see Eq.~\eqref{gamma_th_matrix}). Schematic representation of the equation for the scattering vertex function $\Gamma^{\rm th}$ between two Fermions (blue lines) mediated by a \fpm{mediator} Boson (wiggled line) initially prepared in a Gaussian thermal state. The properties of the Boson are encoded in the Green's function $D$, while the form factor of the Boson-Fermion coupling is $V^{(0)}$. The matrix $M_{\rm th}$ encodes the momentum, frequency, and causal structure of the scattering (with the symbol $\circ$ indicating the tensor product in {all these} spaces). The scattering structure can be directly solved by inverting the corresponding matrix (see Eq.~\eqref{gamma_th_matrix}).} 
   \label{fig:hierarchy_thermal}
   \end{figure*}

 Here we will compute the physical
 $\Gamma$ for \fpm{mediators} initialized in a pure Fock state $\hat{\rho}(0)
 = | \{ n_{s} \} \rangle \langle \{ n_{s} \} |$. In this case the physical vertex
$\Gamma^{\{n_s\}}$ is obtained by substituting Eq.~\eqref{Gamma_ladder_separated} in Eq.~\eqref{Gamma_ph} and then taking the derivatives with respect to the sources:
 \begin{eqnarray}
 \Gamma^{\{n_s\}} = \prod \limits_{s} \frac{1}{n_s!} \left( \frac{ \partial}{\partial u_s} \right)^{n_s} \left[\frac{\Gamma(u) }{\prod \limits_{r} (1-u_r)}\right]\Bigg|_{u_s=0}, \nonumber 
 \end{eqnarray}
 to obtain
 \begin{eqnarray}
 \Gamma^{\{n_s\}} = V^{(0)}{D}+M \circ\Gamma^{\{n_s\}}+
    \!\!\!\!\!\!\!\!\!\sum \limits_{\substack{\{p_s\} \\ \sum \limits_s p_s < \sum \limits_s n_s}}\!\!\!\!\!\!\!\!\! F_{n_s-p_s}  M_{\mathrm{res}} \circ
    \Gamma^{\{p_s\}},~~~~
    \label{eq:hierarchy_general}
 \end{eqnarray}
 with
\[
  F_{n_s-p_s} =\frac{[(\partial/\partial u_s)^{n_s-p_s} f(u_s)]|_{u_s=0}}{(n_s-p_s)!}
\]
where $p_s\le n_s \ \forall s$ and $M=M_{\mathrm{nonres}}+M_{\mathrm{res}}$. \color{black}{$f(u_s)$ is defined below Eq.~\eqref{Gamma_ladder_separated}}.
The physical vertex function for the generic initial density matrix shown in
Eq.~\eqref{rho0_diagonal} can then be constructed as
$\Gamma=\sum_{\{n_s\}} c_{\{  n_{s}\}} \Gamma^{\{n_s\}}$.

Equation \eqref{eq:hierarchy_general} is the main result here. This
new tool provides the desired generalization of the Bethe-Salpeter equation for the
vertex function to the situation where the \fpm{mediator} is
initialized in a Fock state (or an arbitrary mixture of Fock
states). Equation \eqref{eq:hierarchy_general}, schematically
represented in Fig.~\ref{fig:hierarchy} for the single-mode case, clearly
shows how this generalized form of the equation contains an additional
structure with respect to the standard Bethe-Salpeter equation, since it involves
a new set of vertices (or components of the vertex tensor), each
corresponding to a different Fock state. These additional components
satisfy a set of coupled equations of the Dyson type. Since each of
the additional components of the vertex tensor corresponds to a given
Fock state, one can think of this extra structure being a
representation of the Hilbert space of the \fpm{mediator}. This is why
we named the above equation ``\fpm{mediator}-Hilbert-Space Bethe-Salpeter''
(MHSBS) equation.
For a \fpm{mediator} initialized in a given Fock-state, we see that the
MHSBS equation involves a vertex component for that Fock state, plus
an additional vertex component for each of all the Fock states with a
smaller number of Bosons. The corresponding set of coupled Dyson equations can be inverted as
\begin{equation}
 \Gamma^{\{n_s\}}=  (1-M)^{-1} \circ \left[ V^{(0)}{D}+\!\!\!\!\!\!\!\!\sum \limits_{\substack{\{p_s\} \\ \sum \limits_s p_s < \sum \limits_s n_s}} \!\!\!\! F_{n_s-p_s}  M_{\mathrm{res}} \circ  \Gamma^{\{p_s\}} \right] 
 \label{eq:hier_iterative}
\end{equation} 
  and then solved in a iterative manner starting from the vertex
  component corresponding to the Fock vacuum state ($n_s=0$
  $\forall s$),
 \begin{eqnarray}
 \Gamma^{\{0\}} = (1-M)^{-1} \circ \left[ V^{(0)}{D} \right]. 
 \label{Gamma_BS_inverted}
 \end{eqnarray}
{\color{black} We explicitly show the cascading structure of the vertex function when the \fpm{mediator} Boson is initialized in a pure Fock state focusing on the single mode Boson case. The explicit $u$-dependence in Eqs.~\eqref{Gamma_ladder} and \eqref{Gamma_ladder_separated} depends on the details of which component of $\Gamma(u)$ in Keldysh space we are calculating. For illustration purpose (without calculating a complete structure of $\Gamma(u)$ in Keldysh space), we assume here that $\Gamma(u)$ in Eq.~\eqref{Gamma_ladder} contains atleast a term with $D_K(u)$. In this case, $f(u)= (1+u)/(1-u)$ which yields,
\begin{eqnarray}
    F_{n-p} &=& 1, ~\mathrm{for}~n=p \nonumber \\
    &=& 2 , ~\mathrm{for}~0<p\leq (n-1)
\end{eqnarray}
Starting from a vacuum state, we get the higher Fock states iteratively,
\begin{widetext}
\begin{eqnarray}
    \Gamma^{1}&=& \left[ (1-M)^{-1} + F_1  (1-M)^{-1} \circ M_{\mathrm{res}} \circ (1-M)^{-1}    \right] \circ  \left[ V^{(0)}{D} \right] \nonumber \\
    \Gamma^{2}&=& \Big[ (1-M)^{-1} + (F_1+F_2) (1-M)^{-1} \circ M_{\mathrm{res}} \circ (1-M)^{-1}    + \nonumber \\
    && ~~~~4  (1-M)^{-1} \circ M_{\mathrm{res}} \circ (1-M)^{-1} \circ M_{\mathrm{res}} \circ (1-M)^{-1}  
    \Big] \circ  \left[ V^{(0)}{D} \right] \nonumber \\ 
    \vdots&& \nonumber \\
    \Gamma^{n}&=& \Big[ (1-M)^{-1} + \left(\sum \limits_{m=1}^{n} F_m \right)  (1-M)^{-1} \circ M_{\mathrm{res}} \circ (1-M)^{-1}    + \dots \nonumber \\
    && ~~~~F_1^n  (1-M)^{-1} \circ M_{\mathrm{res}} \circ (1-M)^{-1} \circ \dots \circ M_{\mathrm{res}} \circ (1-M)^{-1}  
    \Big] \circ  \left[ V^{(0)}{D} \right] 
    \label{Gamma_n_explicit}
\end{eqnarray}
\end{widetext}
This is pictorially shown in Fig.~\ref{fig:hierarchy}.

\fpm{The above expression has been obtained by considering a Fock state of the mediator, and is otherwise applicable to the broad class of setups where the Fermion-Boson coupling given in Eq.~\eqref{Yukawa} is valid. This comes at the price of it not showing any explicit dependence on external parameters, which is hidden in the form of the matrices $M$ and $M_{\rm res}$ (the function $F_m$ is a numerical constant). Still, it clearly shows that the vertex function $\Gamma_n$ will have a different functional dependence on the external parameters with respect to the case of a Gaussian state of the mediator. Indeed, while in the latter case the matrix $1/(1-M)$ appears linearly in the vertex function (compare Fig.\ref{fig:hierarchy} with Fig.\ref{fig:hierarchy_thermal} and Eq.~\eqref{Gamma_n_explicit} with Eq.~\eqref{gamma_th_matrix} in the next section on the thermal situation), for a Fock state of the mediator with $n$ photons, the matrix $1/(1-M)$ (and in turn all the external parameters) enter through a polynomial of order $n$. This qualitatively affects the two-body scattering, and in turn collective phenomena, as we shall explicitly demonstrate below in Section \ref{critical}.}
}
 
\subsection{Recovering the thermal-equilibrium Bethe-Salpeter equation} \label{thermal_general}
 
In this sub-Section, starting from the generally applicable (MHSBS) equation, we will recover the standard Bethe-Salpeter equation for the two-particle vertex function for the case where the \fpm{mediator} Bosons are in a Gaussian thermal state at equilibrium with the Fermions. This is by far not simply a sanity check for our MHSBS approach, but also physically very insightful. As will shall see next, it shows that the thermal density matrix given in Eq.~\eqref{rho0_thermal} is a special mixture of all the Fock states which allows for a rearrangement of the emergent Hilbert-space structure in the MHSBS equation, such that this additional structure becomes irrelevant.

To illustrate this, we take the example of a fermionic system coupled to a single mode of Bosons where the physical two-particle vertex function $\Gamma^{\mathrm{th}}$ of the system in the thermal case can be constructed from the weighted average of the vertex function of the constituent Fock states $\Gamma^{n}$  as,
 \begin{equation}
 \Gamma^{\mathrm{th}}=\frac{\sum \limits_{n} e^{-\beta \omega n}\Gamma^{n} }{\sum \limits_{n} e^{-\beta \omega n}}.
 \label{gamma_th}
 \end{equation}
 Substituting $\Gamma^{ n}$ from Eq.~\eqref{eq:hierarchy_general}, we get,
  \begin{equation}
 \Gamma^{\mathrm{th}}=V^{(0)}{D}+M \circ\Gamma^{\mathrm{th}}+
    \frac{1}{\sum \limits_{n} e^{-\beta \omega n}}\sum \limits_{n} e^{-\beta \omega n}\sum \limits_{p<n} F_{n-p}  M_{\mathrm{res}} \circ
    \Gamma^{p} .
    \label{gamma_inter_th}
 \end{equation}
 {The last term on the R.H.S., which is the additional structure characterizing the MHSBS equation, can be rearranged as,}
 \begin{widetext}
 \begin{eqnarray}
 \frac{1}{\sum \limits_p e^{-\beta \omega p}} \Bigg[&&\left(\sum \limits_{n=1}^{\infty} e^{-\beta \omega n} F_{n} \right) M_{\mathrm{res}} \circ \Gamma^0 + \left(\sum \limits_{n=2}^{\infty} e^{-\beta \omega n} F_{n-1} \right) M_{\mathrm{res}} \circ \Gamma^1 + \left(\sum \limits_{n=3}^{\infty} e^{-\beta \omega n} F_{n-2} \right) M_{\mathrm{res}} \circ \Gamma^2 + ...\Bigg] \nonumber \\
&=&  \left(\sum \limits_{n=1}^{\infty} e^{-\beta \omega n} F_{n} \right) M_{\mathrm{res}} \circ {\sum \limits_{p=0}^{\infty} \Gamma^p  e^{-\beta \omega p}}\frac{1}{\sum \limits_{p} e^{-\beta \omega p}}
  =\left(\sum \limits_{n=1}^{\infty} e^{-\beta \omega n} F_{n} \right) M_{\mathrm{res}} \circ  \Gamma^{\mathrm{th}},
 \end{eqnarray}
 \end{widetext}
where in the first line we have collected all the terms with fixed $p$, and in the second line we have shifted all the sums over $n$ to start from $n=1$ and we have identified $\Gamma^{\mathrm{th}}$ with Eq.~\eqref{gamma_th}. This means that the MHSBS equation for $\Gamma^{\mathrm{th}}$ given in Eq.~\eqref{gamma_inter_th} {can be directly closed without having to solve the emergent Hilbert-space structure, yielding}
 \begin{eqnarray}
 \Gamma^{\mathrm{th}}&=&V^{(0)}{D}+M_{\mathrm{th}} \circ\Gamma^{\mathrm{th}}, \nonumber \\
 \Rightarrow \Gamma^{\mathrm{th}}&=& (1-M_{\mathrm{th}})^{-1} \circ \left[ V^{(0)}{D} \right]
 \label{gamma_th_matrix}
 \end{eqnarray}
 where $M_{\mathrm{th}}=M+\left(\sum \limits_{n=1}^{\infty} e^{-\beta \omega n} F_{n} \right) M_{\mathrm{res}} $.

 The above line of arguments can be easily  generalized to multi-mode Bosons to recover the structure of the standard BS equation in thermal equilibrium shown in Fig.~\ref{fig:hierarchy_thermal}.

 \section{Superconductivity mediated by cavity photons prepared in non-Gaussian states}
 \label{LIS}

 In the previous Section, we have considered a generic system of
 Fermions interacting with a bosonic field via a Yukawa-type coupling.
 We have developed the MHSBS formalism to non-perturbatively compute
 the Fermion-Fermion scattering vertex for the case where the \fpm{mediator}
 Boson is initialized in an arbitrary quantum state.

In this Section and in the remainder of this manuscript, we will
 focus on a specific example, where a particular non-relativistic version of QED is
 realized by confining light within a material. This is currently experimentally achieved
both with electrons in solid state \cite{garcia2021manipulating,schlawin2021cavity} and with synthetic materials made of ultracold atomic
gases \cite{mivehvar_2021}. The confinement of light (for instance
via mirrors forming a cavity) has two consequences
which make these platforms ideal for exploring the phenomenology of
interest. First, it allows to separate certain modes from the
electromagnetic continuum and selectively couple them to the
Fermions. Second, it enhances the QED coupling
so that the equivalent of the
fine-structure constant can become so large that already low intensity
light (and even electromagnetic vacuum fluctuations) can affect a
finite density of Fermions. Therefore, standard quantum optics
techniques become available for preparing desired non-classical and non-Gaussian states of a small number of photons in few selected
modes, also including spatial modes \cite{rubinsztein-dunlop_roadmap_2016,forbes_quantum_2019,ra_non-gaussian_2020-1}. This allows to explore how \fpm{mediator}-state preparation affects the photon-mediated
interactions and, since Fermions have fixed and controllable density,
also collective phenomena like instabilities toward macroscopically
ordered states.

In general, the above regime of QED within materials is an exciting playground
for the study of collective phenomena, and several directions have
emerged both for the solid-state \cite{schlawin2021cavity} and ultracold-gas
implementation \cite{mivehvar_2021}. While our MHSBS approach is not limited to a certain
class of collective phenomena, we will here consider the case of
photon-mediated pairing and the resulting instability toward a
superconducting phase of the Fermions \cite{Schlawin2019,Gao2020,Colella_2019,Schlawin2019a,sheikan_2019}. Recent works have studied
various properties peculiar to this type of pairing, which is
characterized by the fact that the photon-mediator has a gapped spectrum
(due to the cavity mirrors) and at the same time is, at the relevant energies, very delocalized
over the material or, equivalently, 
localized within a narrow region in momentum space, the latter
being much smaller than the Fermi momentum for the
solid-state realizations, but not necessarily for ultracold
atoms. The peculiar properties revealed in these studies emerged
already within the standard BCS-type approximation, where the photons
adiabatically follow the Fermion dynamics, so that they
effectively mediate a static interaction \cite{Schlawin2019}. In particular, for certain
Hubbard-type models, it has been
shown that the sign of the interaction can be different if the photons are
assumed to be in pure Fock states \cite{li_manipulating_2020}.

In a recent work \cite{Chakraborty}, we showed however that
the spatially delocalized cavity photons can mediate an additional,
non-BCS type of resonant pairing. Here the Fermions fully absorb (or emit) a
photon at its own characteristic cavity frequency. These resonant
pairing processes can be dominant over BCS-type processes and are of a completely different
nature. As we shall show in the next Section, these processes are the
ones that depend on the quantum state of the mediator photon and
generate the additional structure characterizing the MHSBS when the
photon is prepared in non-Gaussian states.

\subsection{Pairing-vertex function mediated by cavity photons} \label{BCSnonBCS}

Here we will derive the MHSBS equation for the vertex function in
the pairing channel and the mediator photon prepared in an arbitrary
initial quantum state. Our model for QED consists of Fermions with
dispersion $\epsilon_k$ (measured from the Fermi surface with energy
$E_F$ and momentum $k_F$), coupled to photons in a single standing-wave
electromagnetic mode of the cavity with wavevector $\vec{q}_0$ and resonance frequency $\delta_c$.
This {allows to express the Yukawa interaction Hamiltonian given in Eq.~\eqref{Yukawa} in the momentum basis as:}
\begin{equation}
\label{H_lighmatter}
H_{\rm light-matter} =\sqrt{2\delta_c}\sum \limits_{ \vec{k},\sigma} \sum \limits_{\vec{q}=\pm q_0 \hat{x}} g_0~ c^{\dagger}_{\vec{k}+\vec{q},\sigma}c_{\vec{k},\sigma} X.
\end{equation}
We note that, while as in the standard QED the foundamental coupling is always
  between photons and the Fermion current (minimal coupling), we are
  considering here a coupling involving the Fermion density. We choose
  this variation because of two reasons. First, it can be experimentally
realized using laser-assisted two-photon transitions both for
electrons in solid state materials \cite{Gao2020,chiocchetta_2021} and
ultracold fermionic atoms \cite{mivehvar_2021}; it also takes the same form
if the photons are replaced by collective degrees of freedom, like lattice
phonons \cite{kittel2021introduction} or Bogoliubov modes in a condensate \cite{cotlet_2016}, as realizable both in
charge tuneable monolayer semiconductors \cite{tan_interacting_2020} or ultracold Bose-Fermi
mixtures \cite{hadzibabic_two-species_2002,schreck_quasipure_2001}.
Second, it simplifies the momentum structure of the Yukawa coupling and thus the analysis of
the vertex function. With these implementations in mind, the photon
characteristic frequency $\delta_c$ is not the absolute value of the cavity
resonance frequency but rather its detuning from the laser
frequency. A further advantage of these implementation is indeed the
tuneability of the Hamiltonian via the laser parameters, not only its
frequency but also the value of the Yukawa coupling constant
$g_0$, which is experimentally tuneable by the intensity of the laser
beam. \fpm{We stress that, despite involving the electron density and not the current, the electromagnetic field quadrature appearing in Eq.\eqref{H_lighmatter} pertains to the cavity vector potential in a two-photon transition where one photon is provided by the laser.}
Finally, the coupling of the Fermions to a single standing-wave
  electromagnetic mode directly applies to the nanoplasmonic split-ring 
  cavities used in the solid-state implementations \cite{Gao2020}, as well as to the
  Fabry-Perot cavities used in the atomic-gas
  implementations \cite{mivehvar_2021}. Moreover,
  as discussed in Ref.~\cite{Chakraborty}, as long as the
  characteristic photon momentum satisfies $q_0 \ll k_F$, which we
  will assume throughout, our treatment of the momentum structure of
  the vertex function does also apply to the coupling of the Fermions
  to a continuum of transverse cavity modes, as realized in
  solid-state microcavities \cite{Schlawin2019} or in the atomic-gas context using confocal cavities \cite{lev_2020}.

For the specific QED model of Eq.~\eqref{H_lighmatter}, we obtain
  the effective action as a sum of terms in Keldysh space with the structure \eqref{S_eff}, describing the Fermion-Fermion
  interactions mediated by the photons, with the momentum-dependent
  coupling taking the simple form $V^{(0)}(\vec{q})= g_0^2 \delta_c
  \delta_{\vec{q},\pm\vec{q}_0} $. 
The equation for the intermediate, $u$-dependent vertex function
\eqref{Gamma_ladder}, now written by making the Keldysh structure
explicit, reads (we refer to the Appendix \ref{app:vertex} for more details on the derivation)
\begin{equation}
  \Gamma(u)=V^{(0)}D_{A}+M_{\rm{BCS}}\circ \Gamma(u)+\frac{1+u}{1-u}M_{\rm{res}}\circ\Gamma(u),
 \label{eq:BS}
\end{equation}
where we defined the matrix-multiplication in momentum-frequency space
as $M\circ \Gamma(u)\equiv 
 \sum_{\vec{k}} \int \frac{d\omega_1}{2\pi}
M(\vec{p}\omega,\vec{k}\omega_1)\Gamma(\vec{k},\omega_1;u)$, and with the matrices
\begin{widetext}
\begin{eqnarray}
M_{\rm{BCS}}(\vec{p}\omega,\vec{k}\omega_1)& = &\mathbf{i}  V^{(0)}(\vec{k}-\vec{p}) \left[{D}_{A}(\omega_1-\omega) ~G_{K}(\vec{k},\omega_1) ~G_{R}(-\vec{k} ,-\omega_1)+{D}_{R}(\omega_1-\omega) ~G_{R}(\vec{k},\omega_1) ~G_{K}(-\vec{k} ,-\omega_1)\right] \nonumber \\
M_{\rm{res}}(\vec{p}\omega,\vec{k}\omega_1) &=& \mathbf{i} V^{(0)}(\vec{k}-\vec{p}) 2 \mathbf{i}\mathrm{Im}\left[{D}_{R}(\omega_1-\omega)\right] ~G_{R}(\vec{k},\omega_1) ~G_{R}(-\vec{k} ,-\omega_1)
\label{BS_RK}
\end{eqnarray}
\end{widetext}
defining the coupling between the different momentum and frequency components of the vertex function. We note that the two-particle vertex function $\Gamma(P,p,p')$ is in general a function of the COM coordinates $P=(\vec{P},\Omega)$, incoming relative coordinates $p=(\vec{p},\omega)$ and the outgoing relative coordinates $p'=(\vec{p}\ ',\omega')$. However, Eq.~\eqref{eq:BS}, written in the pairing channel as per Eq.~\eqref{BS_RK} (corresponding to choosing the two Fermion-lines in
  the Feynman diagram of Fig.~\ref{fig:hierarchy} and \ref{fig:hierarchy_thermal} to be
  co-propagating), is a coupled equation only in the relative incoming coordinates, i.e. $\Gamma(P,p,p')$ couples to $\Gamma(P,k,p')$ and the kernel $M$ within the ladder approximation is independent of $p'$. Moreover, since we want to study the
  superconducting instability we set $P=0$ and express the vertex function in the simpler notation $\Gamma(\vec{p},\omega)=\Gamma(P=0,p,p')$, which describes the scattering of two Fermions with
  equal and opposite momenta and frequency $(\vec{p},\omega)$ and $(-\vec{p},-\omega)$.
Here we restricted our
  calculation only to those initial density matrices of the photons
  which keep the dynamics time-translation invariant (see discussion
  in Section \ref{BSIC}), which allowed us to write the equation \eqref{eq:BS} in frequency space.

The frequency-momentum and causal structure of Eq. \eqref{eq:BS} at thermal equilibrium has been discussed in Ref.~\cite{Chakraborty}, and we limit ourselves to review the main features below, in order to leave space for the discussion of the emergent Hilbert-space structure we focus on here.
Let us first consider the second term on the R.H.S. of Eq.~\eqref{eq:BS}, containing $M_{\rm BCS}$. It comes from the non-resonant processes discussed in the previous section and, as
  expected, it does not carry information about the initial state of
  the \fpm{mediator}. As we shall see next, it leads to the standard BCS scenario. It can be understood as two Fermions interacting via a retarded/advanced
  potential, whose frequency-dependence is set by the
  Fourier transform of the corresponding photon GF given in
  the second line of Eq.~\eqref{DK_u}: $D_{R(A)}(\omega_1-\omega)= 1/[2(\omega_1 -\omega \pm \mathbf{i}0^+)^2-2\delta_c^2]$. Such pairing
  processes thus need the presence of Fermions, whose distribution is
  thermal and encoded in the Fourier transform of the Keldysh GF of
  Eq. \eqref{G_thermal}. 
  For the sake of simplicity of the momentum structure, we start by considering the case of the cavity momentum $q_0\to 0$. In these non-resonant
  processes, Fermions always remain in the vicinity of the Fermi
  surface (FS) with momentum $\vec{p}\sim \vec{k} =\vec{k}_F$. We also assume that the Fermion distribution function is broadened by the finite lifetime of the quasi-particles as 
\begin{equation}
{G}_{K} (\vec{k},\omega_1)=\frac{1}{\omega_1-\epsilon_{\vec{k}} -  \mathbf{i}\tau^{-1}}~ \frac{ \left[ -2 \mathbf{i}\tau^{-1}\tanh \left( \frac{\omega_1}{2T} \right) \right]}{\omega_1-\epsilon_{\vec{k}} +  \mathbf{i}\tau^{-1}}. \nonumber
\end{equation}
This function has a pole in frequency at the characteristics energy scale of the Fermions near the FS $\omega_1 \sim \mathbf{i}\tau^{-1}$ in $M_{\rm BCS}$ of
Eq.~\eqref{BS_RK}. 
Assuming that the photon gap is large: $\delta_c \gg
\tau^{-1}$, and considering only the strongest pairing which
happens in the limit of vanishing incoming relative frequency $\omega$, the frequency dependence of the retarded interaction potential mediated by photons can be neglected, resulting in the usual static attractive
interaction characterizing the BCS scenario. 
The vertex function of the low energy Fermions which form pairs under the attractive interaction defines the on-shell vertex function $\Gamma_{\rm on-shell}(u)=\Gamma(|\vec{p}|\sim k_F, \omega \sim 0;u) \approx \Gamma(|\vec{p}| \sim k_F, \omega \sim \mathbf{i}\tau^{-1};u)$ . 
This identification to a single component  $\Gamma_{\rm on-shell}$ is valid as long as $\tau^{-1} \ll \delta_c$. 
 In the case of finite cavity momentum $q_0\ll k_F$, BCS processes couple the vertex function of the low-energy Fermions in the vicinity of the FS at momenta $|\vec{p}| \sim k_F, k_{F}\pm q_0$ and $\omega \sim 0,\epsilon_{k_F\pm q_0}$. Again, for $\epsilon_{k_F\pm q_0} \ll \delta_c$ 
 we can identify all the components of the vertex
  function within this small energy and momentum shell around the FS
  as a single component  $\Gamma_{\rm on-shell}$, so that $M_{\rm
    BCS}$ becomes a diagonal matrix, coupling $\Gamma_{\rm
    on-shell}(u)$ to itself.

The situation is different for the resonant processes encoded in
  $M_{\rm res}$. The fact that these involve the full absorption/emission of a
  photon is formally expressed by the presence of
  \[
2\mathbf{i}\mathrm{Im}D_{R}(\omega_1-\omega)=\frac{-\pi \mathbf{i}}{2\delta_c} \left[   \delta(\omega_1-\omega - \delta_c) - \delta(\omega_1-\omega + \delta_c) \right],
\]
which is peaked at the cavity resonance frequency. As expected, these are the processes that carry the information about the initial state of
the \fpm{mediator} in Eq.~\eqref{eq:BS}.
These processes are non-diagonal in the
frequency space of the vertex function, since a resonant photon
scatters the incoming on-shell electrons ($\omega \sim
0$) to energies far away from the FS: $\omega_1 \sim \pm \delta_c$.

Using the Fourier transform of the retarded/advanced Fermion GF:
$G_{R(A)} (\vec{k}, \omega)=1/(\omega-\epsilon_{k} \pm \mathbf{i}
\tau^{-1})$, Eq.~\eqref{eq:BS} becomes (see Appendix \ref{app:vertex} for more details)
\begin{equation}
\begin{split}
\Gamma_{\mathrm{on-shell}}(T;u)&=\Gamma_0+ \frac{2 \tilde{g}
  \delta_c}{T} \Gamma_{\mathrm{on-shell}}(u)\\
  -2  \tilde{g} \left( \frac{1+u}{1-u} \right)&  \left[
  \Gamma(\vec{k}_F,\delta_c;u)+
  \Gamma(\vec{k}_F,-\delta_c;u) \right].
  \label{eq:BS_u_freq}
  \end{split}
\end{equation}
Here $\tilde{g}=g_0^2/( 2 \delta_c)^2$ and the bare vertex $\Gamma_0=V^{(0)}(\vec{k}-\vec{p})D_{A}(\omega_1-\omega) \approx -g_0^2/(2\delta_c)$ .

In order to efficiently pair, the off-shell electrons have to
undergo a second scattering process involving a resonant photon which
brings them back to the on-shell region close to the FS. 
To compute this second-order process,
one first uses Eq.~\eqref{eq:BS} to find the first off-shell component of the vertex  $\Gamma(\vec{k}_F,\pm\delta_c;u)$ as a function of the on-shell component only, which is done by neglecting the coupling to higher off-shell components. One then 
substitutes the result for $\Gamma(\vec{k}_F,\pm\delta_c;u)$ into Eq.~\eqref{eq:BS_u_freq} (this truncation is
allowed for large photon gap \cite{Chakraborty}).

This yields the following closed equation for the on-shell
component of the intermediate vertex function:
\begin{widetext}
\begin{eqnarray}
\Gamma_{\mathrm{on-shell}}(T;u)=\Gamma_0+ \frac{2\tilde{g} \delta_c}{T} \Gamma_{\mathrm{on-shell}}(u) 
+ 8 \left( \frac{\tilde{g}\delta_c}{\tau^{-1}} \right)^2 \left( \frac{1+u}{1-u} \right)^2  \Gamma_{\mathrm{on-shell}}(u).~~~~~
\label{Gamma_OS_u}
\end{eqnarray}
\end{widetext}

In the case when the \fpm{mediator} photon is initialized in a thermal Gaussian
  state, the above equation holds directly for the physical
  $u$-independent vertex function, whereby the fraction involving $u$
  appearing in the last term is substituted by the photon distribution
  function (see Eq.~\eqref{eq:Gamma_th_closed}). This thermal situation has
  by itself interesting properties which have been discussed in a
  previous work \cite{Chakraborty}.
 Here we limit ourselves to repeating that the second term on the
 R.H.S. represents the standard BCS contribution, where the polynomial
 dependence on $T$ is due to the {delocalized nature of the photons} i.e.~the delta-peak of the
 momentum-dependent coupling function $V^{(0)}(\vec{p})$. On the other
 hand, the third term on R.H.S. results from the resonant pairing
 processes, which carry the information about the initial \fpm{mediator's} state
 via the source $u$. We note that, although the resonant pairing is a second-order process involving scattering to the off-shell frequency sector and back, the small energy denominator ($\sim \tau^{-1}$) set by the characteristics energy scale of the low energy electrons {can make} it dominate over the BCS processes. {The fact that the repeated absorption and emission of full photons remains resonant is due to the narrow cavity resonance together with the delocalized nature of the photons thanks to which they transfer a well-defined momentum.}
Even though the above equation for the intermediate vertex
  function has a closed form in frequency- and momentum-space, the
  explicit presence of the source $u$ generates an additional structure in
the equation for the physical vertex function. We demonstrated this
for the general case of Section \ref{BSIC}, and in the next Section we
will see this structure emerging for the specific pairing scenario
considered here.

\subsection{Photons prepared in a pure Fock state} \label{Fock_vertex}

We now focus on the case where the \fpm{mediator} photons are initially
prepared in a pure Fock state described by the density matrix
$\hat{\rho}_0=|n \rangle \langle n |$.
This can be achieved with many different protocols, and the most
  suitable one for the present purpose will depend on the specific
  realization of our QED within materials. Here we limit ourself to
  observe that the possible strategies can be divided in two classes:
  i) a suitable light source is used to
inject the photon Fock state inside the cavity, or ii) the photon Fock
state is directly prepared inside the cavity. 
For the first class, sources of non-classical states of structured light are already available and being continuously developed \cite{rubinsztein-dunlop_roadmap_2016,forbes_quantum_2019,ra_non-gaussian_2020-1,erhard2020advances}, the additional step required for our purposes being the in-coupling of the light pulse into the cavity modes to which the matter couples. For the second class, several options exist which rely on coupling the desired cavity modes with some material’s degree of freedom inside the cavity. In the case where our QED model is implemented using atomic gases, the additional degree of freedom can be for instance an internal electronic transition with a laser-assisted coupling to the cavity modes of interest (see e.g.~\cite{hacker_deterministic_2019}). This laser is then to be turned off after the state preparation, then another laser is used to dispersively couple the atomic motion to the cavity modes for the implementation of the QED coupling of Eq.~\eqref{H_lighmatter}. On the other hand, for solid-state implementations of our QED model, a piece of nonlinear medium can be added to the intracavity hetherostructure, to which also the material hosting the quantum matter belongs. The coupling to the nonlinear medium can be then controlled by further driving lasers, as considered for instance in recent promising schemes \cite{lingenfelter_andrew_unconditional_nodate}. An analogue situation can be realized with ultracold atoms, where the additional nonlinearity is brought in by a micromechanical element coupled to the cavity \cite{hunger_resonant_2010,zhong_millikelvin_2017}. 

Our goal is to compute the
corresponding physical on-shell vertex function, which we denote by
$\Gamma^n_{\mathrm{on-shell}}$. Following the procedure illustrated in
Eq.~\eqref{eq:hierarchy_general} for the general case, we derive the
equation for $\Gamma^n_{\mathrm{on-shell}}$ from the intermediate $u-$dependent vertex function $\Gamma_{\mathrm{on-shell}}(u)$ given in
Eq.~\eqref{Gamma_OS_u}. We obtain 
 \begin{widetext}
 \begin{equation}
 \Gamma^n_{\mathrm{on-shell}}(T) = \Gamma_0+ \frac{ 2 \tilde{g} \delta_c}{T} \Gamma^n_{\mathrm{on-shell}}+ 8 \left( \frac{\tilde{g}\delta_c}{\tau^{-1}} \right)^2 \left[  \Gamma^n_{\mathrm{on-shell}} +4\sum \limits_{m=0}^{n-1}(n-m) \Gamma^m_{\mathrm{on-shell}} \right].
  \label{Gamma_os_Fock}
 \end{equation}
  \end{widetext}
  As anticipated, despite the closed-form of the equation for the
  intermediate vertex $\Gamma_{\rm on-shell}(T;u)$, the equation for the
physical vertex $\Gamma^n_{\mathrm{on-shell}}(T)$ is not closed. A new
set of vertex components emerges, reflecting the Hilbert space
structure of the initial quantum state of the photon \fpm{mediator}. As we
have seen also in the general case of Section \ref{BSIC}, each of
these additional components corresponds indeed to a given Fock state,
with a number of photons lower than the initial photon
number. Physically, this implies that the Fermion-Fermion vertex
mediated by the Boson prepared in a pure Fock state with a given
number of photons,
effectively involves scattering processes mediated by all Fock states
with a smaller
numbers of resonant photons, down to the vacuum state. 
These scattering processes take place in different Hilbert space
sectors of the \fpm{mediator} photon, and are organized hierarchically
(as noted already in the general case of Eq.~\eqref{Gamma_BS_inverted}),
which means we can obtain the physical vertex $
\Gamma^{n}_{\mathrm{on-shell}}(T)$ in an iterative way starting from
  the vacuum state. The hierarchy is expressed in terms of the bare
  vertex $\Gamma_0$ and the functions  $N_{\mathrm{vac}}=32(\tilde{g}\delta_c)^2/(\tau^{-1})^2$ and $M_{\mathrm{vac}}(T)=2\tilde{g}
\delta_c/T+8(\tilde{g}\delta_c)^2/(\tau^{-1})^2$. The latter two are combinations of the entries of the matrix $M_{\rm BCS}+M_{\rm res}$, which governs the coupling between the different momentum-frequency components of the vertex.
The iterative solution then reads
\begin{eqnarray}
  \label{gamma_n_itr}
  \Gamma_{\mathrm{on-shell}}^0(T)&=&\frac{\Gamma_0 }{1-M_{\mathrm{vac}}}, \nonumber \\
  \Gamma_{\mathrm{on-shell}}^1(T)&=&\Gamma_0 \left[ \frac{1}{1-M_{\mathrm{vac}}}+  \frac{N_{\mathrm{vac}}}{(1-M_{\mathrm{vac}})^2}  \right], \nonumber \\
  \Gamma_{\mathrm{on-shell}}^2(T)&=&\! \Gamma_0 \left[ \frac{1}{1-M_{\mathrm{vac}}}+  \frac{3 N_{\mathrm{vac}}}{(1-M_{\mathrm{vac}})^2}+ \frac{N_{\mathrm{vac}}^2}{(1-M_{\mathrm{vac}})^3}  \right]~~~~~~ \nonumber \\
  .~~~~~~&&\\
  .~~~~~~&&\nonumber
  \end{eqnarray}

 At this point we can already note that as $n$ increases $ \Gamma^n_{\mathrm{on-shell}}(T)$ shows a stronger divergence near $M_{\mathrm{vac}}(T)=1$,
 which corresponds to the pairing instability point i.e. the critical
 point of the transition towards a superconducting state. This \fpmm{modifies the} critical behaviour near the superconducting transition, when the photons are initialized in different Fock states. This will be discussed in detail in Sec.\ref{critical}.

 \subsection{Photons in thermal equilibrium with electrons}

{As done for the general case in Section \ref{thermal_general}, we will compute here the vertex function for the thermal case starting from the MHSBS equation.} 
When the photons are initialized in a Gaussian thermal state at temperature $T=1/\beta$ equal to the one of the electrons, the on-shell vertex function satisfies the following equation,
 \begin{widetext}
 \begin{eqnarray}
 \Gamma^{\mathrm{th}}_{\mathrm{on-shell}}(T)&=&\frac{1}{\sum \limits_{n=0}^{\infty} e^{-\beta n \delta_c}} \sum \limits_{n=0}^{\infty} e^{-\beta n \delta_c} \Gamma^n_{\mathrm{on-shell}}, \nonumber \\
 &=&\Gamma_0 +2\frac{ \tilde{g}\delta_c}{T}\Gamma^{\mathrm{th}}_{\mathrm{on-shell}} +  8\left( \frac{\tilde{g}\delta_c}{\tau^{-1}}  \right)^2 \left[ \Gamma^{\mathrm{th}}_{\mathrm{on-shell}} + \frac{4}{\sum \limits_{n=0}^{\infty} e^{-\beta n \delta_c}}  \sum \limits_{n= 1}^{\infty} e^{-\beta n \delta_c} \sum \limits_{m=0}^{n-1}(n-m) \Gamma^m_{\mathrm{on-shell}} \right],
 \label{Gamma_th_bc}
 \end{eqnarray}
 where in the last equation we have substituted $\Gamma^n_{\mathrm{on-shell}}(T)$ from Eq.~\eqref{Gamma_os_Fock}. As we discussed in Section \ref{thermal_general}, we can further organize the last term on the R.H.S. of the above equation by collecting coefficients of $\Gamma^m_{\mathrm{on-shell}}(T)$ in the infinite series to obtain a closed-form equation in terms of $ \Gamma^{\mathrm{th}}_{\mathrm{on-shell}}$,
 \begin{equation}
  \Gamma^{\mathrm{th}}_{\mathrm{on-shell}}(T)=  \Gamma_0 +\left[ \frac{ 2 \tilde{g}\delta_c}{T} +  8 \left( \frac{\tilde{g}\delta_c}{\tau^{-1}}  \right)^2  \coth^2 \left( \frac{\delta_c}{2 T}  \right) \right] \Gamma^{\mathrm{th}}_{\mathrm{on-shell}}(T).
\label{eq:Gamma_th_closed}
\end{equation}
\end{widetext}
Instead of the Fock-state hierarchy of equations given in Eq.~\eqref{gamma_n_itr} containing $M_{\rm vac}(T)$ and $N_{\mathrm{vac}}$, we obtain here a single equation containing  $M_{\mathrm{th}}(T)= 2 \tilde{g}\delta_c/T + 8 (\tilde{g}\delta_c/\tau^{-1}  )^2  \coth^2( \delta_c/2 T )$.
As seen in the general case, this is possible since the thermal Gaussian state allows for a rearrangement of the Hilbert-space structure of the scattering such that the the latter becomes irrelevant, and the equation for the on-shell vertex function can be directly closed.

\section{Critical Properties of Superconducting Transition} \label{critical}
In the previous Section, we have computed the two-particle scattering vertex
  function for electrons interacting via photons according to the QED
  model of Eq.~\eqref{H_lighmatter}. Being interested in the
  superconducting phase transition, we computed the scattering in the
  pairing channel.
As we have seen, within the standard BCS approximation, cavity photons mediate a
static, attractive interaction potential, provided that the detuning
of the cavity resonance frequency is positive ($\delta_c >0$). This
attractive interaction is further enhanced by the non-BCS processes
involving absorption and emission of resonant photons owing to the
long-range nature of the interaction. At the
critical temperature for the superconducting transition, the
  vertex function in the pairing channel diverges, indicating an
  instability towards the formation of two-particle bound states,
i.e., Cooper pairs with opposite momentum \cite{abrikosov2012methods}. In this
Section, we will use the MHSBS formalism to study the the critical
properties of the superconducting transition, comparing the case
  where the photon is in a thermally mixed Gaussian state at
  equilibrium with the electrons, with the case where
  it is prepared in a pure Fock state.

With this aim, we introduce the pairing field $\Delta^{P}(p)$ (also
known as superconducting gap field), which is in general a function of
the coordinates of both electrons in the pair, or equivalently of
both the relative $p$ and the COM coordinate $P$ (see Section
\ref{BCSnonBCS}). It can be used as an auxiliary field to decouple
the quartic electron-electron interaction term $S_{\rm eff}$ in
Eq.~\eqref{S_eff} into a form which is diagonal in the the COM coordinate:
\begin{eqnarray}
e^{\mathbf{i}S_{\rm eff}}= \prod \limits_{P}&& \int D[\Delta^{P}(p)] e^{\mathbf{i} \int d{p} ~d{p}' \Delta^{P*}(p) ~\Gamma_0^{-1}(p,p';u)~\Delta^{P}(p')} \times \nonumber \\
&&e^{\mathbf{i} \int dp \Delta^{P}(p) \psi^{*}(P+ p) \psi^*(P- p)+h.c.}.
\label{HS_decoupled_action}
\end{eqnarray} 
Here
$\Gamma_0^{-1}(p,p';u)=1/(V^{(0)}(\vec{p}'-\vec{p})D(\omega'-\omega;u))$. This
procedure, known as Hubbard-Stratonovich decoupling
\cite{kamenevbook,altland}, is a property of Gaussian path integrals.
The resulting action is quadratic in the fermionic fields $\psi$, which
can thus be integrated out exactly to get an action $S(\Delta;u)$
solely in terms of the pairing field. Up to quadratic order, this
action reads
\begin{equation}
e^{\mathbf{i}S_{\rm eff}}=\prod \limits_{P} \int D[\Delta^{P}(p)] e^{\mathbf{i}\int d{p}~ d{p}' ~\Delta^{P*}(p) \Gamma^{-1}(P,p,p';u) \Delta^{P}(p')},
\label{action_delta}
\end{equation}
where $\Gamma(P,p,p';u)$ is the $u-$ dependent two-particle
  vertex function introduced in the previous Section. From the action
  \eqref{action_delta}, we see that it corresponds to the Gaussian GF
of the pairing field $\Delta^{P}(p)$. The physical two particle
vertex function $\Gamma(P,p,p')$ (and thus the physical pair GF) corresponding to a given initial
density matrix $\hat{\rho}_{0}$ of the mediator photon is then
obtained from $\Gamma(P,p,p';u)$ through the proper set of
  $u$-derivatives described in the previous Sections.
Next, we will analyze the pair GF
$\Gamma(P,p,p')$, computed via the MHSBS formalism, at and in the
vicinity of the transition point, in order to extract some of the
critical properties.

 \subsection{Critical temperature} \label{subsec:Tc}

 At the critical point, the low energy electrons having zero COM
 momentum and frequency ($P=0$), as well as relative coordinates at the FS
($|\vec{p}|\approx k_F,\omega\approx0$) become unstable towards pair formation. This is
 manifested as a pole in the on-shell two-particle vertex function
 $\Gamma_{\mathrm{on-shell}}(T)=\Gamma(P=0;|\vec{p}|\approx k_F,\omega\approx0;p')$ at $T=T_c$,
 associated with a uniform pairing field $\Delta^{P=0}(\vec{p}=\vec{k}_F,\omega=0)$. In
 this sub-Section, we will calculate the critical temperature using
 $\Gamma_{\mathrm{on-shell}}(T)$ obtained in Section \ref{LIS}. 

 We first consider the equilibrium case where the photons are initialized
 in a thermal and Gaussian density matrix at a temperature $T$ equal to that of the electrons. In this case, the physical on-shell vertex function $\Gamma^{\mathrm{th}}_{\mathrm{on-shell}}$ given in Eq.~\eqref{eq:Gamma_th_closed} diverges when {$M_{\rm th}(T)=1$, yielding the equation of the critical temperature:}
 \begin{equation}
2\frac{ \tilde{g}\delta_c}{T_c^{\mathrm{th}}} +  8 \left( \frac{\tilde{g}\delta_c}{\tau^{-1}}  \right)^2  \coth^2 \left( \frac{\delta_c}{2 T_c^{\mathrm{th}}}  \right) =1.
\label{Tc_th}
 \end{equation}
We discussed the solution of the above equation for
$T_{c}^{\mathrm{th}}$ in Ref.~\cite{Chakraborty}, considering a
Fermi-liquid type lifetime of electrons due to the screened Coulomb
interaction $\tau^{-1}(T)=(\pi T^2/8 E_F)\log(E_F/T)$. 
\fpm{We note that, in principle, the presence of a phase transition and the associated critical boson fluctuations can strongly affect the quasiparticle lifetime and eventually lead to non-Fermi-liquid behavior. The computation of this effect in our case is more challenging than in the standard non-Fermi-liquid literature (see \cite{rao_2023} for an example involving electrons in cavities), since the interaction mediator (photon) is not the boson that becomes critical at the superconducting phase transition. As explained in the previous section, the critical boson in our case is the Cooper pair, whose propagator is essentially the vertex function $\Gamma$. Therefore, in order to compute the effect of critical fluctuations onto the electrons, our case would require a self-consistent T-matrix approach. Due to the tensorial structure of the vertex function, such approaches become quite demanding, so that the literature so far is essentially restricted to static contact interactions (see e.g. \cite{pini_2024} for a study of the associated non-Fermi-liquid behavior), allowing to bypass the tensorial structure of $\Gamma$ thanks to the absence of an actual phonon or photon mediator.}
Coming back to our case, for realistic parameters in 2D materials (STO/LAO) coupled to a tera-Hertz
split-ring cavity, the superconducting transition in the thermal case
occurs in the low Kelvin regime. We note that the photon resonance
frequency in these cavities satisfies $\delta_c \gg T_c$, so that the
average number of photons in the cavity is negligible: $\langle
\hat{n}\rangle= 0.5\coth \left( \frac{\delta_c}{2 T_c^{\mathrm{th}}}
\right)-0.5 \sim 0$. The critical temperature in this vacuum case
can be calculated from,
\begin{equation}
T_c^{\mathrm{vac}}=\frac{ 2 \tilde{g}\delta_c}{1- 8 \left( \frac{\tilde{g}\delta_c}{\tau^{-1}}  \right)^2  }.
\label{eq:Tc_vac}
\end{equation}  
{It is however important to note, that the actual critical temperature in the thermal equilibrium case, though very close to $T_{\rm vac}$, is always going to be slightly larger due to the average number of photons not being exactly zero.}

 We now turn our attention to the case of photons prepared in a Fock
 state, $\hat{\rho}_0=|n\rangle \langle n|$. The corresponding
 on-shell physical vertex function $\Gamma^n_{\mathrm{on-shell}}$
 satisfies Eq.~\eqref{Gamma_os_Fock}. As evident from the iterative
 solution for $\Gamma^n_{\mathrm{on-shell}}$ given in
 Eq.~\eqref{gamma_n_itr}, for any finite photon number $n$ in the Fock
 state, $\Gamma^n_{\mathrm{on-shell}}$ diverges when the denominator
 $M_{\mathrm{vac}}=2 \tilde{g} \delta_c/T+  8
(\tilde{g}\delta_c)^2/(\tau^{-1})^2$ approaches to $1$. This implies
 that the superconducting transition occurs at the same critical
 temperature $T_c^{\mathrm{vac}}$ as the thermal vacuum case,
 independently of the number of photons in the Fock state.

\subsection{Susceptibility exponent} \label{subsec:susceptibility}
Since $\Gamma$ is equivalent to the propagator of the
  pairing field, the exponent characterizing the divergence of
  $\Gamma_{\mathrm{on-shell}}(T-T_c)$ as $T$ approaches $T_c$
  corresponds to the critical susceptibility exponent $\gamma$:
\begin{equation}
\Gamma_{\mathrm{on-shell}}(T-T_c) \sim \frac{1}{|T-T_c|^{\gamma}}.
\label{eq:def_suscep}
\end{equation}

We again first consider a Gaussian thermal initial state of the
mediator photons. In this case,
$\Gamma_{\mathrm{on-shell}}^{\mathrm{th}}$ given in
Eq.\eqref{eq:Gamma_th_closed} diverges when
$M_{\mathrm{th}}(T=T_c^{\mathrm{th}})=1$ (see Eq.\eqref{Tc_th}). In the
regime of interest $T\ll E_F$, $M_{\mathrm{th}}(T)= 2 \tilde{g}\delta_c/T +  8(\tilde{g}\delta_c/\tau^{-1}  )^2  \coth^2(
\delta_c/2 T )$ (including the temperature dependent life-time of the
quasi-particles) is a smooth function of $T$. Hence, in order to
analyze the nature of the singularity in
$\Gamma_{\mathrm{on-shell}}^{\mathrm{th}}$, we can expand
$M_{\mathrm{th}}(T)$ in a Taylor's series around $T=T_c^{\mathrm{th}}$, to obtain
\begin{equation}
\Gamma_{\mathrm{on-shell}}^{\mathrm{th}} \propto \frac{\Gamma_0}{T-T_c^{\mathrm{th}}}.
\label{Gam_T_th}
\end{equation}
This yields $\gamma=1$, which recovers the standard BCS result of the
susceptibility critical exponent for the Gaussian fixed point in thermal equilibrium.
That is, when the mediator photon is initialized in a Gaussian thermal
state, the superconducting transition temperature is strongly modified by the
non-BCS resonant pairing processes (see Section \ref{BCSnonBCS}), but the
critical behaviour of the pairing susceptibility remains unaltered from the standard BCS case.

We now turn to the case where the mediator photon is initialized in a
Fock state. For Fock states with any finite number of
photons $n$, $\Gamma_{\mathrm{on-shell}}^n$ given in
Eq.~\eqref{Gamma_os_Fock} always diverges at the vacuum critical
temperature $T_c^{\mathrm{vac}}$, defined by
$M_{\mathrm{vac}}(T=T_c^{\mathrm{vac}})=1$. As apparent from Eq.~\eqref{gamma_n_itr}, in the vicinity of the phase transition, the strongest divergence in $\Gamma_{\mathrm{on-shell}}^n
$ comes in the form,
\begin{equation}\label{eq:n-dep_div_gamma}
\Gamma_{\mathrm{on-shell}}^n \sim \frac{N_{\mathrm{vac}}^{n}\,\Gamma_0}{\left(1-M_{\mathrm{vac}}(T)\right)^{n+1}} \propto \frac{N_{\mathrm{vac}}^{n} \,\Gamma_0}{\left(T-T_c^{\mathrm{vac}}\right)^{n+1}},
\end{equation}
where we expanded $M_{\mathrm{vac}}$ in Taylor's series around the
critical temperature $T_c^{\mathrm{vac}}$. This yields the
susceptibility exponent $\gamma=n+1$. That is, while the critical
  temperature remains unaffected, the critical behaviour is dramatically altered, as
  the critical exponent becomes linearly proportional to the number of photons prepared in
  the Fock state.

\subsection{Correlation-length exponent}
\label{subsec:correlation_length}
Finally, we study the critical behaviour of the spatial correlations of the order parameter. To do so, we define the function $\tilde{\Delta}(\vec{P})$ as the pairing field $\Delta^{P}(p)$ introduced in Eq.~\eqref{HS_decoupled_action} evaluated at $P=(\vec{P},\Omega=0)$ and $p=(\vec{p}=\vec{k}_F,\omega=0)$, i.e.~with the variables $\Omega$, $\vec{p}$ and $\omega$ fixed to the value at which they give the most divergent contribution to the pair GF $\Gamma(\vec{P},\Omega;\vec{p},\omega;p')$ when approaching the phase transition. We can then define the correlation function:
\begin{equation}
\begin{split}
\langle \tilde{\Delta}(\vec{r}_1) \tilde{\Delta}^*(\vec{r}_2)\rangle  &=  \int  \frac{d\vec{P}}{(2\pi)^d} \langle \,\tilde{\Delta}(\vec{P}) \tilde{\Delta}^{*}(\vec{P})\rangle  e^{\mathbf{i}\vec{P}\cdot(\vec{r}_1-\vec{r}_2)} \\
= \int  \frac{d\vec{P}}{(2\pi)^d} &\Gamma(\vec{P},\Omega=0;\vec{p}=\vec{k}_F,\omega=0;p') \, e^{\mathbf{i}\vec{P}\cdot(\vec{r}_1-\vec{r}_2)}.
\end{split}
\end{equation}
In order to obtain the
long-distance behaviour of the correlation function with spatial
separation $|\vec{r}_1-\vec{r}_2|$, we expand
$\Gamma(\vec{P},\Omega=0;\vec{p}=\vec{k}_F,\omega=0;p')$ in
Taylor's series around $\vec{P}= 0$ and $T=T_c$ to the lowest
non-trivial order. Near the critical point, the spatial correlation decays exponentially $\sim \exp(-|\vec{r}_1-\vec{r}_2|/\xi(T))$ with a characteristic correlation length $\xi(T)\sim (T-T_c)^{-\nu}$. $\xi(T)$ diverges at $T=T_c$ with the critical exponent $\nu$ and the system becomes scale-invariant at the phase transition.

Again, we first consider the mediator photons being prepared in a
Gaussian thermal state. In this case, the physical two-particle vertex
function
$\Gamma^{\mathrm{th}}(\vec{P},\Omega=0 ;\vec{p}=\vec{k}_F,\omega=0;p')=\tilde{\Gamma}^{\mathrm{th}}(\vec{P},T)$,
expanded around $\vec{P}=0$ and $T=T_c^{\mathrm{th}}$, reads (to leading order)
\begin{equation}
\tilde{\Gamma}^{\mathrm{th}}(\vec{P},T) = \frac{\Gamma_0}{{P}^2\left( \cos^2\theta +a  \right)+m}.
\end{equation}
Here $\theta$ is the angle between the COM momentum $\vec{P}$ and the relative momentum of the incoming electrons
$\vec{p}=\vec{k}_F$ and $a$ is a constant (independent of $\vec{P}$
and $T$) depending on the characteristic energy scales of the
electrons and photons at $T=T_c^{\mathrm{th}}$. $m=m(T-T_c^{\mathrm{th}})\propto (T-T_c^{\mathrm{th}})$ is the
mass gap obtained by expanding $M_{\mathrm{th}}(T)$ upto linear order around $T=T_c^{\mathrm{th}}$ (see Eq.~\eqref{Gam_T_th}).
The above form of $\tilde{\Gamma}^{\mathrm{th}}(\vec{P},T)$ can be
obtained analytically, by Taylor expansion of $\tilde{\Gamma}^{\mathrm{th}}(\vec{P},T)$ around $\vec{P}=0$ and $T=T_c^{\mathrm{th}}$. Here we
motivate this form by general symmetry arguments. 
In the scattering events
contributing to $\tilde{\Gamma}^{\mathrm{th}}(\vec{P},T)$, two
incoming electrons close to the FS and having momentum
$\vec{P}+\vec{k}_F$ and $\vec{P}-\vec{k}_F$, interact by exchanging
cavity photons of momentum $\vec{q}_0$ and are scattered to the momenta $\vec{P}+\vec{k}_F+\vec{q}_0$ and $\vec{P}-\vec{k}_F-\vec{q}_0$ respectively near the
FS. Hence, as a consequence of the delta-like shape of the momentum structure of the cavity photons $V^{(0)}(\vec{q})\propto
  \delta_{\vec{q},\pm\vec{q}_0} $, the relative momentum of the scattered electrons is fixed. This breaks rotational symmetry of  the scattering amplitude and hence
$\tilde{\Gamma}^{\mathrm{th}}(\vec{P},T)$ depends on $\theta$ through $\epsilon_{\vec{P}\pm\vec{k}_F}$ for $q_0 \approx 0$
(unlike the standard case of phonon-mediated superconductivity). Still, inversion symmetry $\vec{P}\rightarrow -\vec{P}$ prohibits terms linear in $\vec{P}$ (and hence in $\cos\theta$) to appear in the expansion of $\tilde{\Gamma}^{\mathrm{th}}(\vec{P},T)$.

To obtain the spatial correlation function, we then perform the
Fourier transform,
\begin{equation}
\int \frac{d\vec{P}}{(2\pi)^d} \tilde{\Gamma}^{\mathrm{th}}(\vec{P},T)~e^{\mathbf{i}\vec{P} \cdot (\vec{r}_1-\vec{r}_2)} \propto e^{-|\vec{r}_1-\vec{r}_2|\sqrt{\frac{m}{1+a}}}.
\end{equation}
This implies that the correlation length $\xi$ diverges as,
\begin{equation}
\xi \propto   \frac{1}{\sqrt{m}} \propto \frac{1}{(T-T_c^{\mathrm{th}})^{\frac{1}{2}}},
\end{equation}
that is, $\nu=\frac{1}{2}$. This correlation length exponent for
photons prepared in a Gaussian thermal state is the same as in
the standard BCS scenario.

We now finally consider photons initialized in a Fock state. The spatial behaviour of the correlation function is extracted
from the physical vertex function
$\Gamma^n(\vec{P},\Omega=0;\vec{p}=\vec{k}_F,\omega=0;p')=\tilde{\Gamma}^n(\vec{P},T)$
expanded the critical point $\vec{P}=0,T=T_c^{\mathrm{vac}}$. As for the
thermal case, the functional dependence of
$\tilde{\Gamma}^n(\vec{P},T)$ on ${P}^2$ and $\theta$ is constrained
by the symmetries of the problem, while the leading order temperature
dependence comes from the expansion of $M_{\mathrm{vac}}(T)$ around
$T=T_c^{\mathrm{vac}}$ through the temperature dependent 
  mass gap $m=m(T-T_c^{\mathrm{vac}})$. However, the crucial difference from the thermal
case lies in the polynomially stronger divergence of the
$\Gamma$ at the critical point, leading to the following form
\begin{eqnarray}
\tilde{\Gamma}^{n}(\vec{P},T) &=& \frac{\Gamma_0}{\left[{P}^2\left( \cos^2\theta +a  \right)+m  \right]^{n+1}}, \nonumber \\
& \approx & \frac{1}{m^n(n+1)} \frac{\Gamma_0}{\left[{P}^2\left( \cos^2\theta +a  \right)+\frac{m}{n+1}\right]},~~~
\end{eqnarray}
where in the last equation we expanded the denominator in a binomial
series and kept the lowest-order terms in $P$. We see that mass gap
$m$ is now effectively reduced by a factor $n+1$, yielding an exponential decay of the spatial correlations of the form
\begin{equation}
\int \frac{d\vec{P}}{(2\pi)^d}\tilde{\Gamma}^n(\vec{P},T) ~ e^{\mathbf{i}\vec{P} \cdot (\vec{r}_1-\vec{r}_2)}\propto e^{-|\vec{r}_1-\vec{r}_2|\sqrt{\frac{m}{(1+a)(n+1)}}}.
\end{equation}
Hence, the correlation length reads
\[
  \xi \propto \left(\frac{n+1}{T-T_c}\right)^{1/2}.
\]
We thus see that, while the correlation-length exponent remains
  unaltered from the BCS case $\nu=1/2$, the correlation length scales
with the square root of the number of photons in the Fock state. This
result completes the derivation of the critical properties summarized
in Fig.~\ref{table:result}.

\fpm{\subsection{\fpmm{Discussion} of the change in \fpmm{the critical exponents}}}

\fpmm{Within our formalism, the critical exponents start to depend on the number of mediator-photons in the Fock state. This can be traced back to the additional dimension emerging in the two-body scattering, and corresponding to the Fock space of the mediator photon. Indeed, as clearly shown by our MHSBS equation (see for instance Fig. 2), it is the nontrivial structure in this extended space that, once solved, leads to the photon-number-dependent critical exponents.}

\fpm{The critical exponents are in a way the most difficult thing to change about a phase transition, because of the concept of universality: Even very different models can show the same critical exponents (but dramatically different critical temperatures) as long as they share symmetries and dimensionality. In this spirit, our results thus show the deepest possible change of the critical behavior, and that this can be controlled by the photon number. Actually, due to the very concept of universality, one should be surprised that this is possible at all in our case, since we are not changing symmetries (the order parameter is the same as in the standard BCS approach) and also not spatial dimensionality.}
\fpmm{An interpretation of this surprising finding is that, through the emergent Hilbert-space structure, we are actually changing the effective dimensionality of the space in which two-body scattering takes place, and thereby the critical exponents.}

\fpmm{These findings shall motivate further theoretical investigations focusing on the computation of critical exponents with dedicated approaches, which will require the extension of methods like the renormalization group analysis to the present case featuring the additional Hilbert-space structure of the two-body scattering. Such studies shall eventually address possible modifications of the universality class, as well as the robustness of the photon-number-dependence of the critical exponents.}

\section{Polarizability and the time-independent approximation} \label{sec:polz}

So far, we have considered a time-independent situation in which the
\fpm{mediator} photon remains in its initial quantum state. 
However, the latter is in principle affected by the interactions with the fermionic matter. In
general, if our QED system of Fermions and photons would be ergodic,
the whole system would thermalize and the memory of the initial photon
state would be lost. In fact, the constraints imposed by gauge
invariance can lead to ergodicity breaking through what is called
disorder-free many-body localization \cite{smith_2017,brenes_2018}.
For our QED within materials,
one can thus expect the memory of the initial state never to be lost
completely. How this happens and what consequences it has for the
present phenomenology is a very interesting fundamental question, that
we defer however to future investigations.

Here we instead consider a different mechanism for the preservation of
the memory of the initial photon state, which is simply the strong
suppression of the effect of the material onto the photons \cite{Piazza2014Quantum}.
 
This effect is described by the polarization function, $\Pi(\omega,q_0)$, where
 $\omega$ and $q_0$ are the energy and momentum of the photon. In our
 Keldysh formalism, the polarization function has two components: the retarded,
 $\Pi_R(\omega,q_0)$ and the Keldysh, $\Pi_K(\omega,q_0)$.
 The real part of $\Pi_R(\omega,q_0)$ modifies the dispersion of the
 photons, while its imaginary part, resulting from on-shell processes,
 introduces absorption.
 When the
 fermionic matter is in thermal equilibrium, the imaginary part of the
 polarization function fixes also the Keldysh component via a
 fluctuation dissipation relation: $\Pi_K(\omega,q_0)=2\mathbf{i}
 \coth(\omega/ 2T) \rm{Im}[\Pi_{R}(\omega, q_0)]$. The Keldysh component
 is the relevant one here, as we are 
 interested in the possible modification of the initial quantum state
 of the photons, which we have seen being encoded in the Keldysh GF.
 
As noted already above, a characteristic feature of the regime of QED considered here is the strong separation of energy and momentum scales
between light and matter: compared to the fermionic scales, the cavity
photons have a large energy gap $\delta_c \gg
   \epsilon_{k_F+q_0}(\approx v_F q_0)$ but at the same time very
small momentum $q_0\ll k_F$. Here $v_F$ is the Fermi velocity.

   In the present model of QED, given in Eq.~\eqref{H_lighmatter}, the photons couple to
   the electron density, which is a conserved quantity. This fact, as we argue
   next, together with the above separation of scales, implies that
   the polarization function has to be small, thereby guaranteeing the
   preservation of the memory of the initial quantum state of the photon.
 \begin{figure}[t!]
   \centering
  \includegraphics[width=\columnwidth]{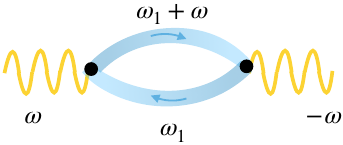}
  \caption{Feynman diagram corresponding to the polarization function
    of the Fermions. It quantifies the backaction of the matter onto
    the light, and in particular the modification of the initial
    quantum state of the photon.} 
   \label{fig:highorder_polarization}
   \end{figure}

   As shown in Ref.~\cite{wardidentity} for a general class of Yukawa
   couplings between a real Boson and Fermions, of which our
   QED coupling is a particular case, this result applies
   every time the interaction form factor is constant. Let us briefly repeat
   the general argument here. Let us first consider the zero-momentum
   case. $\Pi_R(\omega,0)$ is the two-time correlator of the total
   number of electrons $n_f=\sum_{\vec{k},\sigma}
   \psi^{\dagger}_{\vec{k},\sigma}\psi_{\vec{k},\sigma}$, the latter
   being conserved. The time-independence of $n_f$ then results in
   $\Pi_R(\omega,0)=0$. By continuity one can then imply that
   $\Pi_R(\omega,q)$ has to remain small in the limit $\omega\gg v_F
   q$. This limit is exactly the limit of interest in our case where
   $\delta_c\gg v_F q_0$. We thus can conclude that the polarization
   function, whose leading-order diagrammatic expression in shown in
   Fig.~\ref{fig:highorder_polarization}, should be small compared to
   $\delta_c$ even when fully dressed with high-order
   corrections. In Ref.~\cite{wardidentity}, this general
   argument has been verified both perturbatively and non-perturbatively.
 
We finally note that, in the regime of interest here where the polarization function remains small, phase transitions of the superradiant type, characterized by a macroscopic occupation of the cavity mode and a density modulation of the Fermions \cite{Piazza2014Umklapp,Keeling2014fermionic,Chen2014Superradiance,sandner2015self,fraser_2019,Mazza2019,polini_2019,nataf_2019,lev_2020,mora_2020}, are excluded.

\section{Conclusions}
\label{sec:conclusions}

In this work, we developed a field-theoretical framework for the study of the effect of \fpm{mediator}-quantum-state preparation on scattering and collective phenomena. We derived an equation for the two-particle scattering vertex for the case where the force Boson mediating the scattering is initially prepared in an arbitrary quantum state. This MHSBS equation is characterized by an additional structure in an emergent scattering space which adds to the standard energy-momentum space. This emergent scattering structure reflects the wave-function of the quantum state of the force \fpm{mediator} in Hilbert space, and is nontrivial when the latter is not Gaussian. Gaussian and in particular thermal Gaussian states of the mediator instead allow for a reorganization of the additional Hilbert-space structure that renders the latter irrelevant, yielding the standard Bethe-Salpeter equation for the scattering vertex.
As a first example, we applied our MHSBS equation to the case of photon-mediated pairing and superconductivity for electrons in cavities, and showed that photons prepared in pure Fock states strongly enhance pair correlations and ultimately modify superconducting critical properties like the susceptibility exponent or the correlation length, which become dependent on the number of photons.

This work shows that by preparing the quantum state of a \fpm{interaction-mediator} the properties of scattering and the resulting collective phenomena can be deeply altered. By developing a theoretical tool for the systematic investigation of this possibility, it paves the way for further investigations exploring new scenarios in few- and many-body physics. Several different research directions emerge naturally from the present work. Also the study of the effect of other non-Gaussian non-classical states of the \fpm{mediator}, beyond the Fock state on which we have focused here. Or the investigation of different ordering phenomena besides the pairing considered here, and how these can be manipulated by \fpm{mediator}-state preparation, including also (frustrated) magnetism with photon-mediated interactions \cite{Mivehvar2019Cavity,chiocchetta_2021}. QED within materials also allows to address our central question from the complementary perspective where the electrons play the role of the \fpm{mediators} of interactions between photons or polaritons \cite{cotlet_2016,tan_interacting_2020}, for which a non-equilibrium field theory has been recently developed \cite{lang_nonequilibrium_2020,lang_interaction-induced_2020,wasak_zeno_2021,wasak_fermi_2021} that could be extended in the future to include the effect of the \fpm{mediator}-state preparation. 

\fpm{Finally, we note that the possibility for experimental investigation also exists at a less fundamental level, where the role of photons can be played by collective degrees of freedom like phonons \cite{chu_creation_2018, satzinger_quantum_2018}, or by excitons(-polaritons) \cite{tan_interacting_2020,cotlet_2016}. 
}
\fpm{While quantum-state-engineering is currently more developed for photons than for the rest of the candidate \fpm{mediators} like phonons, new promising platforms are emerging. Strong candidates are Bose-Fermi mixtures, like the ones realized in charge-tuneable semiconductors \cite{tan_interacting_2020} or in ultracold atoms \cite{schreck_quasipure_2001,hadzibabic_two-species_2002}. For instance, the bosonic component of an ultracold atomic mixtures can be prepared in a non-classical state using the tuneable interatomic interactions (e.g. Mott phases have been achieved already a long time ago \cite{bloch2008many}). Moreover, a major current endeavour [\cite{wiese2013ultracold,zohar2015quantum,aidelsburger2022cold}] in this field (with important experimental steps already taken) is the implementation of (lattice) gauge theories, where the role of the gauge field is played by one component of the mixture, that can be eventually prepared in a interesting quantum state (e.g. a Mott insulator).}

\begin{acknowledgments}
AC acknowledges Abrahams Postdoctoral Fellowship by Rutgers Center for Material Theory where last part of the work is performed.
\end{acknowledgments}

\appendix

\section{Complete expression of the effective action}\label{app:action}

In the main text, we formulated the MHSBS formalism to study Fermion-Fermion scattering mediated by exchange of Bosons which are prepared in arbitrary non-thermal initial density matrices.  As we discussed in section \ref{sec:background}, the information about the initial conditions of the non-interacting system is incorporated by introducing a set of quadratic sources $\{u_{s}\}$ coupled to the bi-linears of the initial fields and taking appropriate set of derivatives w.r.t.~$\{u_{s}\}$ to compute physical correlation functions (or more generally physical partition function). 

\begin{widetext}
We considered the Fermion-Boson coupling to be a Yukawa-type interaction given by $S_{\mathrm{int}}$ in Eq.~\eqref{S_lightmatter}. Since the action $S_B+S_{\mathrm{int}}$ is a quadratic action in Bosonic fields, we can integrate out the Bosonic fields $X_{cl}$ and $X_Q$ exactly to obtain an effective Fermion-Fermion action $S_{\mathrm{eff}}$ of the form (see Eq.~\eqref{S_eff} of the main text),
\begin{eqnarray}
&&S_{\mathrm{eff}}(u)=- \int \int dt ~dt'_1 \sum \limits_{ \{k_i\},s} V^{(0)}(\{k_i\},s) \nonumber \\
&&\Bigg[ D_{R}(s,t,t') \Big[ \psi^*_{cl} (k_1,t) \big\{ \psi^*_{cl} (k_2,t') \psi_{cl}(k_3,t') +cl \leftrightarrow q \big\} \psi_{q} (k_4,t) \nonumber \\
&&~~~~~~~~~~~~~~~~~~~~~~~~~~~~~~~~~~~~~~~~+
 \psi^*_{q} (k_1,t) \big\{ \psi^*_{cl} (k_2,t') \psi_{cl}(k_3,t') +cl \leftrightarrow q \big\} \psi_{cl} (k_4,t) 
\Big] \nonumber \\
&&
+
D_{A}(s,t,t') \Big[ \psi^*_{cl} (k_1,t) \big\{ \psi^*_{cl} (k_2,t') \psi_{q}(k_3,t') +cl \leftrightarrow q \big\} \psi_{cl} (k_4,t) \nonumber \\
&&~~~~~~~~~~~~~~~~~~~~~~~~~~~~~~~~~~~~~~~~+
 \psi^*_{q} (k_1,t) \big\{ \psi^*_{cl} (k_2,t') \psi_{q}(k_3,t') +cl \leftrightarrow q \big\} \psi_{q} (k_4,t) 
\Big] 
\nonumber \\
&&
+
D_{K}(s,t,t';u) \Big[ \psi^*_{cl} (k_1,t) \big\{ \psi^*_{cl} (k_2,t') \psi_{q}(k_3,t') +cl \leftrightarrow q \big\} \psi_{q} (k_4,t) \nonumber \\
&&~~~~~~~~~~~~~~~~~~~~~~~~~~~~~~~~~~~~~~~~+
 \psi^*_{q} (k_1,t) \big\{ \psi^*_{cl} (k_2,t') \psi_{q}(k_3,t') +cl \leftrightarrow q \big\} \psi_{cl} (k_4,t) 
\Big] 
\Bigg].
\label{app:seff}
\end{eqnarray}
Here the four-Fermion interaction is mediated by the Bosonic Green's functions $D(s,t,t';u)$ where the Keldysh component $D_{K}(s,t,t';u)$ solely carries the initial source dependence in the effective Fermionic theory.

The vertex functions corresponding to the $12$ bare vertices shown in
Eq.~\eqref{app:seff}  obey 12 equations which are coupled in
Keldysh space. As discussed in Ref.~\cite{Chakraborty}, for the cavity-QED example given in
section \ref{BCSnonBCS}, one of those 12 equations decouples from
  the others and is the relevant one to be considered.

\section{Details of the derivation of pairing-vertex function mediated
by cavity photons}\label{app:vertex}

This equation is
written in  Eq.~\eqref{eq:BS} and \eqref{BS_RK} of the main text,
  and reads (in explicit form)
\begin{eqnarray}
&&\Gamma(\vec{P},\Omega;\vec{p},\omega;\vec{p}~',\omega';u)= V^{(0)}(\vec{p}'-\vec{p}) D^{A}(\omega'-\omega) \nonumber \\
&& + \mathbf{i} \sum_{\vec{k}} \int \frac{d\omega_1}{2\pi}  V^{(0)}(\vec{k}-\vec{p}) {D}_{A}(\omega_1-\omega) ~G_{K}(\vec{P}+\vec{k},\Omega+\omega_1) ~G_{R}(\vec{P}-\vec{k} ,\Omega-\omega_1) \Gamma(\vec{P},\Omega;\vec{k},\omega_1;\vec{p}~',\omega';u) \nonumber \\
&& + \mathbf{i} \sum_{\vec{k}} \int  \frac{d\omega_1}{2\pi}  V^{(0)}(\vec{k}-\vec{p}) {D}_{R}(\omega_1-\omega) ~G_{R}(\vec{P}+\vec{k},\Omega+\omega_1) ~G_{K}(\vec{P}-\vec{k} ,\Omega-\omega_1) \Gamma(\vec{P},\Omega;\vec{k},\omega_1;\vec{p}~',\omega';u)\nonumber \\
&& + \mathbf{i} \sum_{\vec{k}} \int \frac{d\omega_1}{2\pi}  V^{(0)}(\vec{k}-\vec{p}) {D}_{K}(\omega_1-\omega;u) ~G_{R}(\vec{P}+\vec{k},\Omega+\omega_1) ~G_{R}(\vec{P}-\vec{k} ,\Omega-\omega_1) \Gamma(\vec{P},\Omega;\vec{k},\omega_1;\vec{p}~',\omega';u) .
\label{app:BS_coupled} 
\end{eqnarray}
We recall that the Green's functions for the fermions are given by $G_{R (A)}(\vec{k},\omega)=1/(\omega-\epsilon_{k}\pm \mathbf{i}\tau^{-1})$ and $G_{K}(\vec{k},\omega)= \tanh (\frac{\omega}{2T}) [G_{R}(\vec{k},\omega)-G_{A}(\vec{k},\omega)]$, while the Green's functions for the bosons are given by $D_{R (A)}(\omega)=1/[2((\omega \pm \mathbf{i}0^{+})^2-\delta_{c}^2)]$ and $D_{K}(\omega,u)=-\frac{\mathbf{i}\pi}{2\delta_{c}} (\frac{1+u}{1-u}) [ \delta(\omega-\delta_{c})-\delta(\omega+\delta_{c})]$. First, we will calculate the 2nd and the 3rd term of the R.H.S. of
Eq.~\eqref{app:BS_coupled} (referred as $M_{\mathrm{BCS}}\circ
\Gamma(u)$  in the main text) which lead to the standard BCS pairing
processes involving adiabatic photons and on-shell electrons. As we
discussed in the main text, in order to probe the superconducting
instability we set $\vec{P}=0,\Omega=0$ and $\vec{p}=\vec{k}_F,
\omega=0$. Also for the sake of simplicity, we assume the cavity
momentum $q_0 \to 0$. This immediately simplifies the structure of the
loop momentum 
sum by replacing it with a factor of $2$ (coming from the two $\pm q_0 \to 0$ terms) and just setting $\vec{k}=\vec{p}=\vec{k}_F$. The loop frequency integral over
$\omega_1$ is performed by the method of contour integration by
choosing the pole of $G_K(\vec{k}_F,\omega_1)$ at $\omega_1=\pm
\mathbf{i}{\tau}^{-1}$. Hence the photonic retarded/advanced Green's
functions are evaluated at frequencies $\omega_1-\omega = \pm
\mathbf{i}{\tau}^{-1} $ which is much smaller than photon resonance
frequency $\delta_c$. This is similar to the standard BCS picture
where the Boson mediates a static attractive potential. 
This yields (for $2$d electrons coupled to cavity photons),
\begin{eqnarray}
\sum_{\vec{k}} \int \frac{d\omega_1}{2\pi} M_{\mathrm{BCS}}(\vec{p}=\vec{k}_F, \omega=0;\vec{k},\omega_1)\Gamma(\vec{k},\omega_1;u) =  2 \tilde{g}\delta_c \! \! \! \sum \limits_{\omega_1=\pm \mathbf{i}\tau^{-1}} \tanh\frac{\omega_1}{2T} \frac{1}{\omega_1} \Gamma(\vec{k}_F,\omega_1;u)= \frac{2 \tilde{g}\delta_c }{T} \Gamma_{\mathrm{on-shell}}(u),
\end{eqnarray}
where in the last equality we have assumed $\tau^{-1} \ll T$.

Next, we will evaluate the resonant contribution given by the 4th term in Eq.~\eqref{app:BS_coupled} on R.H.S.~(referred as $(\frac{1+u}{1-u})M_{\mathrm{res}}\circ
\Gamma(u)$ in the main text). In the case, the delta functions in $D_K(\omega_1-\omega;u) $ set the loop frequency variable $\omega_1 = \pm \delta_c$ to yield,
\begin{eqnarray}
\sum_{\vec{k}} \int \frac{d\omega_1}{2\pi}  M_{\mathrm{res}}(\vec{p}=\vec{k}_F, \omega=0;\vec{k},\omega_1)\Gamma(\vec{k},\omega_1;u) &=& 2 \tilde{g}\delta_c^2 \sum \limits_{\omega_1=\pm \delta_c} G_R(\vec{k}_F, \omega_1) G_R(-\vec{k}_F, -\omega_1) \Gamma(\vec{k}_F,\omega_1;u) \nonumber \\
&=&-  2  \tilde{g}\Bigg[ \Gamma(\vec{k}_F,\delta_c;u)+\Gamma(\vec{k}_F,-\delta_c;u) \Bigg].
\label{app:BS_nclosed}
\end{eqnarray}
The above term makes the equation non-diagonal in frequency space, coupling $\Gamma_{\mathrm{on-shell}}$ to the high-frequency component of the vertex function involving off-shell electrons scattered by resonant photons ($\omega_1 -\omega = \pm \delta_c$).

Now we need to evaluate the BCS components of the equation at the external frequency $\omega=\pm \delta_c$. This yields,
\begin{eqnarray}
 \sum_{\vec{k}} \int \frac{d\omega_1}{2\pi}  M_{\mathrm{BCS}}(\vec{p}=\vec{k}_F, \omega=\pm\delta_c;\vec{k},\omega_1) \Gamma(\vec{k},\omega_1;u)
\approx- \frac{\tilde{g}\delta_c^3}{T} \sum \limits_{\omega_1=\pm \mathbf{i}\tau^{-1}} \left[ \frac{1}{\pm2\mathbf{i}\delta_c\tau^{-1}} \right] \Gamma(\vec{k}_F,\omega_1;u).
\end{eqnarray}
Hence,
\begin{equation}
\sum_{\vec{k}} \int \frac{d\omega_1}{2\pi} \left[M_{\mathrm{BCS}}(\vec{p}=\vec{k}_F,
\omega=\delta_c;\vec{k},\omega_1)+M_{\mathrm{BCS}}(\vec{p}=\vec{k}_F,
\omega=-\delta_c;\vec{k},\omega_1) \right] \Gamma(\vec{k},\omega_1;u) \approx 0.
\end{equation}
The off-shell vertex
  component is thus given by (neglecting the contributions from
  further off-shell sectors like $ \Gamma(\vec{k}_F,\pm 2 \delta_c;u)$
  \cite{Chakraborty})
\begin{equation}
\sum_{\vec{k}} \int  \frac{d\omega_1}{2\pi} M_{\mathrm{res}}(\vec{k}_F, \pm\delta_c;\vec{k},\omega_1)\Gamma(\vec{k},\omega_1;u) = 2\tilde{g}\delta_c^2 G_R(\vec{k}_F,0)G_R(-\vec{k}_F,0) \Gamma(\vec{k}_F,0;u) \approx -\frac{2\tilde{g}\delta_c^2}{\tau^{-2}} \Gamma_{\mathrm{on-shell}}(u).
\end{equation}
Substituting this in Eq.~\eqref{app:BS_nclosed}, we obtain the
closed-form equation for $\Gamma_{\mathrm{on-shell}}(u)$ given in Eq. \eqref{Gamma_OS_u} of the main text.

 \fpm{\section{Vertex Correction of the electron-photon interaction } }
 \begin{figure}[t]
   \centering
  \includegraphics[width=0.8\textwidth]{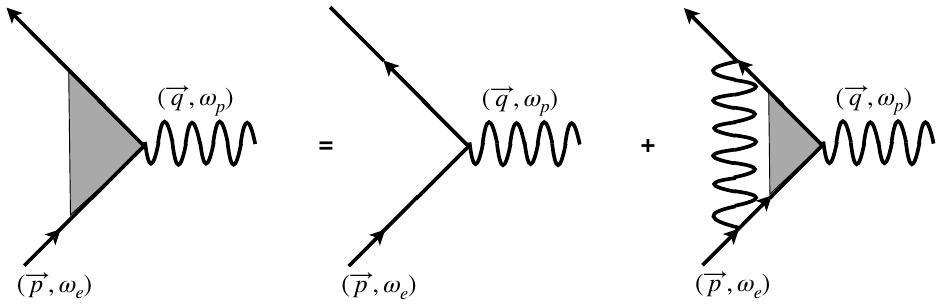}
  \caption{\fpm{This figure shows the fully renormalised electron-photon vertex within ladder approximation. This gives a self-consistent equation for the vertex corrections which is coupled in electronic frequency and momentum. We estimate the non-perturbative vertex correction using the Ward identity for the density vertex.}} 
   \label{fig:ladder_vertex}
   \end{figure}
In this section, we will discuss the effect of the vertex correction
of the electron-photon interaction, which comes from the higher order
diagrams that we have neglected in earlier discussions. 

We will compute the vertex correction $ V_{e-ph}$ non-perturbatively. A widely used choice is to calculate $ V_{e-ph}$
  from the ladder-resummed electron-photon vertex as shown in
  Fig. \ref{fig:ladder_vertex}. The self-consistent equation for the
  renormalized vertex is a coupled equation in the electronic
  coordinates ($\vec{p},\omega_{e}$), the solution of which is in general a nontrivial
  numerical task.   However, some general results can be obtained by using Ward
  identities for the Yukawa vertex \cite{Peskin,wardidentity}. 
We exploit again the fact that the photon momentum $q$ is by far the
smallest scale. This allows us to consider the vertex evaluated at
vanishing photon momentum $\vec{q}\to 0$, for which the following Ward
identity holds \cite{wardidentity} (we suppress the Keldysh structure for the following discussion),
\begin{equation}
1+V_{e-ph} (\vec{p},\omega_{e};\vec{q}\to 0,\omega_p) \propto \frac{1}{\omega_p}\left[ G^{-1}((\vec{p}+\vec{q},\omega_{e}+\omega_p)-  G^{-1}((\vec{p},\omega_{e}) \right].
\label{supp:vertex_Ward}
\end{equation}
Here, $G^{-1}$ is the inverse of the fully dressed electron Green's
function. Within the structure of the Fermi-liquid theory, the
electronic quasi-particles acquire a finite lifetime $\tau$.

We are first interested in the the resonant pairing processes, which carry the information about the
initial \fpm{mediator} state,  mediated by
photons having frequency, $\omega_p \sim \delta_c \gg
\tau^{-1}$. This is the dominant pairing process described in third term in Eq.~\ref{Gamma_OS_u}. In this case Eq. \ref{supp:vertex_Ward} yields a
vertex correction for electrons at the Fermi surface which is given by
\begin{equation}
V_{e-ph} (p\to k_F,\omega_{e}\to 0;\vec{q}\to 0,\omega_p\to \delta_c) \sim  \mathcal{O} \left( \frac{\tau^{-1}}{\delta_c} \right).
\end{equation}
Since we are working in a regime where the decay rate of the
quasi-particles is still much smaller than the photon
frequency $\delta_c$, the vertex correction to the non-BCS term can be
safely neglected.
\\
On the other hand, we note that for the standard BCS mechanism, denoted by second term in Eq.~\ref{Gamma_OS_u},
mediated by adiabatic photons $\omega_p\sim \tau^{-1}$, the numerator of R.H.S of Eq. \ref{supp:vertex_Ward} vanishes for vanishing photon momentum $q=0$. However, for small but finite photon momentum $q_0 \ll k_F$, $V_{e-ph}$ can be finite for the non-resonant BCS process. 

We thus conclude that the peculiarities of cavity photons which
lead to weak polarization effects, namely large frequency and small momentum, also imply that the fully resummed
vertex correction to the electron-photon interaction is small for the resonant pairing, as we have argued using a Ward
identity.

\section{Explicit one-loop calculation of the polarization function}

In this section we provide an explicit calculation (at the perturbative one-loop level) of the polarization function. This calculation is aimed at supporting the general argument for the smallness of polarization effects discussed in section \ref{sec:polz} of the main text.
 
We compute $\Pi_R(\omega,\vec{q}_0)$ perturbatively at the one loop order by evaluating the diagram shown in Fig. \ref{fig:highorder_polarization}. This yields \cite{kamenevbook} (suppressing the momentum-index of the polarization function for compactness),
\beq
\Pi_R(\omega)=\frac{ \mathbf{i}}{2} g_0^2 \delta_c \int \frac{d\vec{k}}{(2\pi)^2} \frac{d\omega}{2\pi} \sum_{\sigma} \left[ G_R(\vec{k}+\vec{q}_0,\omega_1+\omega) G_K(\vec{k},\omega_1) + G_K(\vec{k}+\vec{q}_0,\omega_1+\omega) G_A(\vec{k},\omega_1)  \right],
\eeq
 where the momentum of the external photon line, $\vec{q}_0=\pm q_0 \hat{x}$. After performing the frequency integral, $\Pi_R(\omega)$ takes the form,
 \beq
 \Pi_R(\omega)={2} g_0^2 \delta_c \int \frac{d\vec{k}}{(2\pi)^2} \frac{n_F(\epsilon_{k})-n_F(\epsilon_{|\vec{k}+\vec{q}_0|})}{\omega +\epsilon_{k}-\epsilon_{|\vec{k}+\vec{q}_0|}+\mathbf{i}0^+},
 \label{supp:polarization}
 \eeq
 where, $n_F(\epsilon_{k})=1/\left[\exp(\epsilon_{k}/T)+1 \right]$ is the Fermi function.
 Now, in the limit $T \ll E_F$ we can approximate $[n_F(\epsilon_{k})-n_F(\epsilon_{|\vec{k}+\vec{q}_0|})] \approx \large[\epsilon_{k}-\epsilon_{|\vec{k}+\vec{q}_0|}\large ] \delta(\epsilon_{k})$, which indicates that the energy integral in $\Pi_R(\omega)$ is sharply concentrated around the FS. Moreover, since $q_0 \ll k_F$, we can write $\big[ \epsilon_{k}-\epsilon_{|\vec{k}+\vec{q}_0|} \big] \approx v_k q_0 \cos \theta$, where $v_k=|\vec{k}|/m^*$ and $\theta$ is angle between $\vec{k}$ and $\vec{q}_0$. Inserting these in Eq. \ref{supp:polarization}, we obtain,
 \beq
  \Pi_R(\omega)= \frac{2 g_0^2 \delta_c m^*}{(2\pi)^2} \int d\theta \frac{(-v_F q_0 \cos \theta)}{\omega + \mathbf{i}0^+ -v_F q_0 \cos \theta}
 \eeq
 Separating the real and imaginary part of the above equation, we get 
 \bqa
 \text{Re} \left[ \Pi_R(\omega)\right]&=& \frac{2 g_0^2 \delta_c m^*}{(2\pi)^2} \int  \limits_{0}^{2\pi} d\theta~ P\left[ \frac{(- v_F q_0 \cos \theta)}{\omega  -v_F q_0 \cos \theta} \right] \nonumber \\
 &=& g_0^2 \delta_c \rho (0) \left[ 1-\frac{1}{\sqrt{1- \left( \frac{v_F q_0}{\omega} \right)^2}}  \right]~~~~~~~(\mathrm{for}~ v_F q_0/ \omega<1)~, \nonumber \\
 \text{Im} \left[ \Pi_R(\omega) \right ] &=&  \frac{2 g_0^2 \delta_c m^* \pi}{(2\pi)^2}  \int \limits_{0}^{2\pi} d\theta ~ (v_F q_0 \cos \theta) ~\delta (\omega-v_F q_0 \cos \theta),
 \eqa
where $\rho (0)$ is the density of states of the 2D electrons at the FS. 
 Now, if we analyze the polarization function near the resonance frequency of the on-shell non-adiabatic photons, i.e $\omega \sim \delta_c$, we can expand the denominator of $\text{Re} [ \Pi_R(\omega)]$ in powers of the dimensionless quantity $v_F q_0 / \omega = (E_F / \delta_c) (2q_0 /k_F) \ll 1$ (since $q_0 /k_F \sim 0.002$). The leading order contribution of $\text{Re} [ \Pi_R(\omega)]$ and $\text{Im}[ \Pi_R(\omega)]$ is given by (comparing with the typical energy scale of the photons),
 \beq
  \frac{\text{Re} \left[ \Pi_R(\omega\sim \delta_c) \right ] }{\delta_c^2} \approx -(4\pi)^2 \tilde{g} \delta_c \rho(0) \frac{q_0^2}{k_F^2} \ll 1,~~\mathrm{and},~~\text{Im} \left[ \Pi_R(\omega\sim \delta_c) \right ] =0.
 \eeq
While the imaginary part is zero within this perturbative calculation whenever $\omega>v_f q_0$, in the real part we see the suppression factor $(q_0/k_F)^2$, in accordance with the general argument given in the main text. In principle, there can be and enhancement with the number of electrons, as the density of states at the Fermi surface $\rho(0)$ typically scales with the size of the Fermi surface itself (not actually in a 2D electron gas).
As a matter of fact however, for a standard Fermi-liquid coupled to a Fabry-Perot cavity, the polarization function is typically very small, as discussed e.g. in \cite{andolina2020}. This can be changed by significantly reducing the volume of the electromagnetic modes (see e.g. \cite{todorov2014,helmrich2024ultrastrong}).

\section{\textcolor{blue}{Crossed diagrams of the vertex function}}
 \begin{figure}[t]
   \centering
  \includegraphics[width=1.0\textwidth]{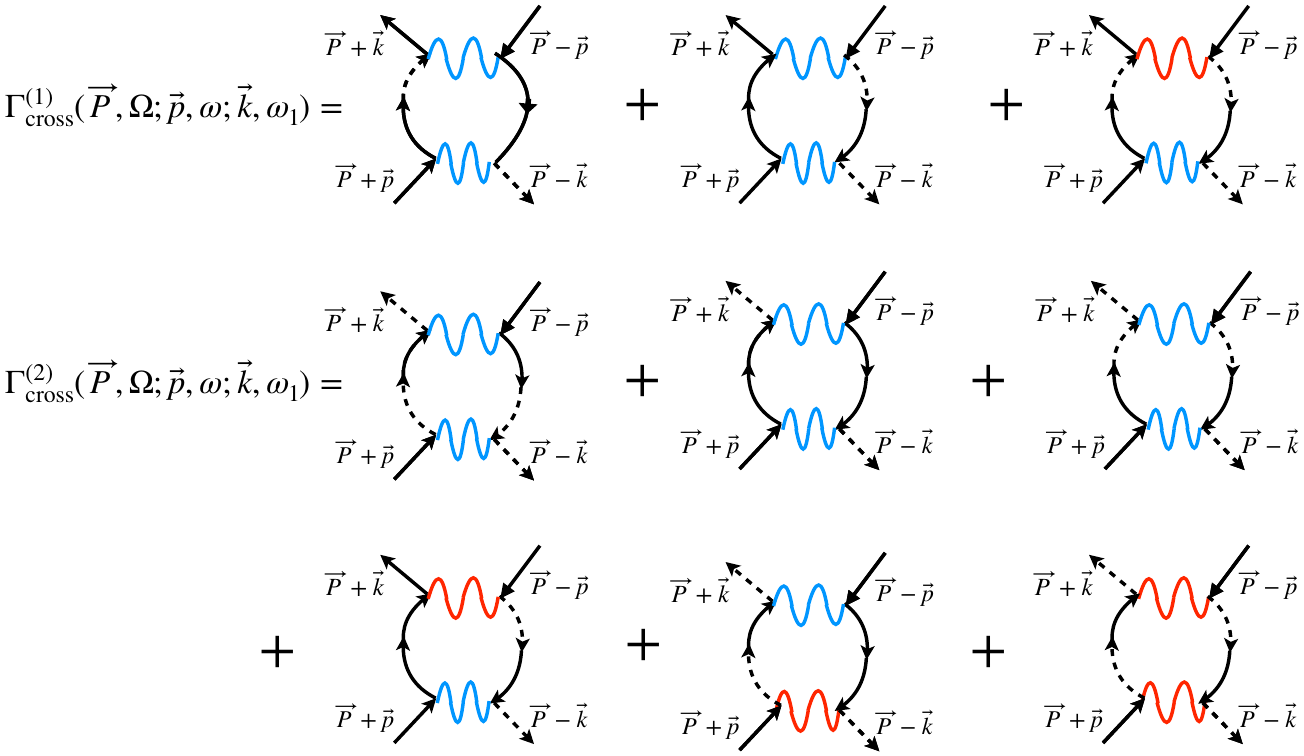}
  \caption{This figure shows the crossed diagrams which lead to corrections of the bare electron-electron pairing vertex.   } 
   \label{fig:highorder_cross}
   \end{figure}

 The general structure of the Bethe-Salpeter equation \cite{abrikosov2012methods} which is used to calculate the transition temperature of the photon-mediated superconductivity is given by (see Eq. \ref{app:BS_coupled} for the full Keldysh structure),
 \bqa
 \Gamma(\vec{P},\Omega;\vec{p},\omega;\vec{p}~',\omega')&=& \tilde{\Gamma}(\vec{P},\Omega;\vec{p},\omega;\vec{p}~',\omega') \nonumber \\
&+& \mathbf{i} \int \frac{d\vec{k}}{(2\pi)^2}  \frac{d\omega_1}{2\pi}   \tilde{\Gamma}(\vec{P},\Omega;\vec{p},\omega;\vec{k},\omega_1)  ~G(\vec{P}+\vec{k},\Omega+\omega_1) ~G(\vec{P}-\vec{k} ,\Omega-\omega_1) \Gamma(\vec{P},\Omega;\vec{k},\omega_1;\vec{p}~',\omega').~
\label{supp:BS_cross}
 \eqa
 Here, $ \tilde{\Gamma}(\vec{P},\Omega;\vec{p},\omega;\vec{p}~',\omega') $ is the set of all irreducible diagrams which can not be divided into two parts by cutting two co-propagating electron lines \cite{abrikosov2012methods}. In our earlier discussion (see Eq. \eqref{app:BS_coupled} and Sec. \ref{BCSnonBCS}), we have approximated $\tilde{\Gamma}$ by $\Gamma_0=V^{(0)}D$ which is the first order diagram in the perturbation series of $\tilde{\Gamma}$. In this section, we will discuss the next higher order diagrams (crossed diagrams) in the perturbation series of $\tilde{\Gamma}$ as shown in Fig. \ref{fig:highorder_cross} . 
 
 The correction to $\tilde{\Gamma}$ coming from the cross diagrams associated with the BCS vertex (i.e corrections to the bare vertex $V^{(0)}D_{A/R}$ in 1st line of Eq.\eqref{BS_RK}) are of the form,
 \bqa
 \Gamma_{\mathrm{cross}}^{(1)}(\vec{P} = 0,\Omega =0;\vec{p},\omega;\vec{k},\omega_1) &=& \mathbf{i} (g_0^2 \delta_c)^2 \int \frac{d\omega_2}{2\pi} \int \frac{d\vec{k}_2}{(2\pi)^2}  \delta_{-\vec{p}-\vec{k}_2,\pm q_0 \hat{x}} \delta_{-\vec{k}-\vec{k}_2,\pm q_0 \hat{x}} \times \nonumber \\
&& \Bigg[4G_{A}(\vec{p}+\vec{k}+\vec{k}_2,\omega+\omega_1+\omega_2) G_K(\vec{k}_2,\omega_2)D_R(-\omega-\omega_2)D_R(-\omega_1-\omega_2)  \nonumber \\
&& +4 G_{K}(\vec{p}+\vec{k}+\vec{k}_2,\omega+\omega_1+\omega_2) G_R(\vec{k}_2,\omega_2)D_A(-\omega-\omega_2)D_R(-\omega_1-\omega_2) + \nonumber \\
&&2 G_{A}(\vec{p}+\vec{k}+\vec{k}_2,\omega+\omega_1+\omega_2) G_R(\vec{k}_2,\omega_2)D_K(-\omega-\omega_2)D_R(-\omega_1-\omega_2) 
 \Bigg],
 \eqa
 where we have set the COM coordinates $\Omega=0$ and $\vec{P}=0$ as discussed in Sec. \ref{BCSnonBCS}. For our present discussion, we will restrict the discussion to the case of thermal photons in equilibrium with electrons. For the sake of simplicity of calculation, we will assume the momentum of the photon propagator $\vec{q}=\pm q_0 \hat{x}$ and show that the above correction terms from cross-diagrams are negligible as long as the separation of energy scales holds, $ \epsilon_{k_F \pm q_0} \ll \delta_c$. In the limit of vanishingly small cavity momentum ($q_0 \rightarrow 0$), the following argument still remains valid as long as the energy scale of the quasi-particles close to FS, $\tau^{-1}\ll \delta_c$. After performing the frequency and the momentum integrals we obtain,
  \bqa
 \Gamma_{\mathrm{cross}}^{(1)}&&(\vec{P} =0,\Omega =0;\vec{p},\omega;\vec{k},\omega_1) =\frac{(g_0^2 \delta_c)^2}{(2\pi)^2}  \left[ \delta_{\vec{k}, \vec{p}} +\delta_{\vec{k},\vec{p}\pm 2 q_0 \hat{x}} \right] \times \sum \limits_{\vec{k}_2=-\vec{p}\mp  q_0\hat{x}} \nonumber \\
&&  \Bigg[ 4  G_{A}(\vec{p}+\vec{k}+\vec{k}_2,\omega+\omega_1+ \epsilon_{|\vec{k}_2|}) D_R(-\omega-\epsilon_{|\vec{k}_2|})D_R(-\omega_1-\epsilon_{|\vec{k}_2|})  \tanh \left ( \frac{ \epsilon_{|\vec{k}_2|}}{2T} \right)  \nonumber \\
&& + 4 G_R(\vec{k}_2, \epsilon_{|\vec{p}+\vec{k}+\vec{k}_2|}-\omega-\omega_1)D_A(-\epsilon_{|\vec{p}+\vec{k}+\vec{k}_2|}+\omega_1)D_R(-\epsilon_{|\vec{p}+\vec{p}'+\vec{k}|}+\omega)  \tanh \left ( \frac{ \epsilon_{|\vec{p}+\vec{k}+\vec{k}_2|}}{2T} \right) + \nonumber \\
&&\! \! \frac{1}{2\delta_c}\coth \left ( \frac{\delta_c}{2T} \right) \left \{ G_{A}(\vec{p}+\vec{k}+\vec{k}_2,\omega_1-\delta_c) G_R(\vec{k}_2,-\omega-\delta_c)D_R(-\omega_1+\omega+\delta_c) +\delta_c \rightarrow -\delta_c \right \}
 \Bigg],~~~~~
 \label{supp:Gamma_cross}
 \eqa
 Here, we will also assume the coordinates of the incoming electrons, $\vec{p}=k_F \hat{x}$ and $\omega \sim \epsilon_{k_F \pm q_0}$.

 First, we will analyze the effect of the corrections introduced by $\Gamma_{\mathrm{cross}}^{(1)}$ on resonant type of pairing discussed in Sec. \ref{BCSnonBCS} (see Eq. \eqref{BS_RK}). As we have discussed in section Sec. \ref{BCSnonBCS}, in the case of non-BCS pairing mechanism, the electrons are scattered far off from the FS, i.e the loop-frequency is close to the photon resonance frequency $\omega_1 \sim \pm \delta_c$ in Eq. \ref{supp:BS_cross}. The corrections coming from $\Gamma_{\mathrm{cross}}^{(1)}(\vec{P}=0,\Omega=0;\vec{p}=k_F\hat{x},\omega\sim \epsilon_{k_F+q_0};\vec{k},\omega_1) $ will contribute in this type of pairing when it has pole w.r.t the variable $\omega_1$ near $\omega_1\sim \pm \delta_c$. However, it can be seen from Eq. \ref{supp:Gamma_cross} that all poles of $\Gamma_{\mathrm{cross}}$ w.r.t $\omega_1$ are in the same half-plane, i.e. in this case upper half of the complex frequency plane. Hence, while performing the loop-frequency integral in the BS equation ( Eq. \ref{supp:BS_cross}), we can always choose the contours in the other half-plane and the poles of  $\Gamma_{\mathrm{cross}}$ w.r.t the loop-frequency will not contribute. This concludes that the corrections coming from the crossed diagrams $\Gamma_{\mathrm{cross}}^{(1)}$ at second order will not affect the non-BCS pairing mechanism.

 	Next, we will evaluate the correction $\Gamma_{\mathrm{cross}}^{(1)}$ introduces to the standard BCS paring when the loop frequency $\omega_1 \sim \epsilon_{k_F\pm q_0}$, i.e the electrons are scattered close to the FS. Compared to the bare vertex $\Gamma_0=-g_0^2/ (2\delta_c)$ of the standard BCS coupling (see below Eq. \eqref{eq:BS_u_freq}), $\Gamma_{\mathrm{cross}}^{(1)}$ obtained from Eq. \ref{supp:Gamma_cross}, introduces a correction term whose relative strength is $\mathcal{O}(\tilde{g}\delta_c/\mathrm{max}(T,\epsilon_{k_F+q_0}))$. 
 	At this point, it is important to note that the new resonant type of pairing proposed in Sec.\ref{BCSnonBCS} increases $T_c$ by more than one order of magnitude (see Ref. \cite{Chakraborty}) and hence further suppresses the corrections coming from $\Gamma_{\mathrm{cross}}^{(1)}$. At typical temperatures of our interest $T\sim 1K$, this correction term coming from $\Gamma_{\mathrm{cross}}^{(1)}$ can be neglected compared to the bare vertex. Once again this discussion illuminates why it is important to include the effects of long-range photon fluctuations mediated by the on-shell non-adiabatic photons in the dynamics. 
 	

Now we calculate the correction to $\tilde{\Gamma}$ coming from the cross diagrams associated with the non-BCS vertex which are of the form,
 \bqa
 \Gamma_{\mathrm{cross}}^{(2)}(\vec{P}=0,\Omega =0;\vec{p},\omega;\vec{k},\omega_1) &=& \mathbf{i} (g_0^2 \delta_c)^2 \int \frac{d\omega_2}{2\pi} \int \frac{d\vec{k}_2}{(2\pi)^2}  \delta_{-\vec{p}-\vec{k}_2,\pm q_0 \hat{x}} \delta_{-\vec{k}-\vec{k}_2,\pm q_0 \hat{x}} \times \nonumber \\
&&4 \Bigg[G_{R}(\vec{p}+\vec{k}+\vec{k}_2,\omega+\omega_1+\omega_2) G_A(\vec{k}_2,\omega_2)D_R(-\omega-\omega_2)D_A(-\omega_1-\omega_2) + \nonumber \\
&& G_{K}(\vec{p}+\vec{k}+\vec{k}_2,\omega+\omega_1+\omega_2) G_K(\vec{k}_2,\omega_2)D_R(-\omega-\omega_2)D_R(-\omega_1-\omega_2) + \nonumber \\
&&  G_{A}(\vec{p}+\vec{k}+\vec{k}_2,\omega+\omega_1+\omega_2) G_R(\vec{k}_2,\omega_2)D_A(-\omega-\omega_2)D_R(-\omega_1-\omega_2) + \nonumber \\
&&  G_{K}(\vec{p}+\vec{k}+\vec{k}_2,\omega+\omega_1+\omega_2) G_R(\vec{k}_2,\omega_2)D_K(-\omega-\omega_2)D_R(-\omega_1-\omega_2) + \nonumber \\
&&  G_{R}(\vec{p}+\vec{k}+\vec{k}_2,\omega+\omega_1+\omega_2) G_K(\vec{k}_2,\omega_2)D_R(-\omega-\omega_2)D_K(-\omega_1-\omega_2) + \nonumber \\
&& G_{R}(\vec{p}+\vec{k}+\vec{k}_2,\omega+\omega_1+\omega_2) G_R(\vec{k}_2,\omega_2)D_K(-\omega-\omega_2)D_K(-\omega_1-\omega_2) 
 \Bigg],
 \label{gamma_cross2}
 \eqa
 Following the line of arguments described above, we will analyze the contribution of $\Gamma_{\mathrm{cross}}^{(2)}$ in correcting the bare non-BCS electron-electron pairing vertex. This correction will come from the poles of $\Gamma_{\mathrm{cross}}^{(2)}$ w.r.t $\omega_1$ which are close to photon resonance frequency $\omega_1 \sim \pm \delta_c$. For this purpose, the relevant terms in $\Gamma_{\mathrm{cross}}^{(2)}$ are those which have poles $\omega_1 \sim \pm \delta_c$ present in both the half planes in the complex frequency plane of $\omega_1$ (similar to the structure of what we have in the bare non-BCS vertex $V_0 D_K(\omega_1-\omega)$). After performing the frequency and the momentum integration in Eq. \ref{gamma_cross2}, we obtain that compared to the bare non-BCS vertex $V_0 D_K(\omega_1-\omega)$, the relative strength of these correction terms are $\mathcal{O}(\tilde{g})$ and $\mathcal{O}(\tilde{g})\tanh(\epsilon_{k_F+q_0}/(2T))$. This verifies that the relative magnitude of these correction terms are much smaller compared to $1$ in the regime of our interest and hence these higher order corrections do not change the main results proposed in the main text. This concludes our discussion on the correction coming from the crossed diagrams in electron-electron pairing vertices. We note that our discussion above is different from the case of a singular gauge propagator discussed in Ref.\onlinecite{AltshulerPRB} (or in Ref.\onlinecite{DanielLossPRB} for finite range interaction).

\end{widetext}

\bibliographystyle{apsrev4-1}
\bibliography{SC_Qlight.bib}

\end{document}